\documentclass[twocolumn, doipub]{aastex631}

\usepackage{graphicx} % Required for inserting images
\usepackage{xcolor}
\usepackage{hhline}
\usepackage{layouts}
\usepackage{ulem}
\usepackage{amsmath}
\usepackage{hyperref}

\newcommand{\sci}[1]{\ensuremath{ \times 10^{#1} }}
\newcommand{\masshalo}{\ensuremath{M_{\mathrm{halo}}}}
\newcommand{\cross}{\ensuremath{ \mathrm{CO}\times \mathrm{CIB}\; }}
\setcitestyle{notesep={ }}      % remove the comma before a citp postnote

\newcommand\numberthis{\addtocounter{equation}{1}\tag{\theequation}}

\begin{document}

% Set Strikethrough Color
\newcommand\ysout{\bgroup\markoverwith{\textcolor{red}{\rule[0.5ex]{2pt}{0.4pt}}}\ULon}

\title{The Modeling Landscape of Extragalactic CO in CMB Surveys}
\author{Yogesh Mehta}
\affiliation{School of Earth and Space Exploration, Arizona State University, 781 Terrace Mall, Tempe, AZ 85287, U.S.A.}

\author{Anirban Roy}
\affiliation{Department of Physics, New York University, 726 Broadway, New York, NY, 10003, USA}
\affiliation{Center for Computational Astrophysics, Flatiron Institute, New York, NY 10010, USA}

\author{Simon Foreman}
\affiliation{Department of Physics, Arizona State University, 550 E.\ Tyler Mall, Tempe, AZ 85287, USA}

\author{Alexander van Engelen}
\affiliation{School of Earth and Space Exploration, Arizona State University, 781 Terrace Mall, Tempe, AZ 85287, U.S.A.}

\author{Nick Battaglia}
\affiliation{Department of Astronomy, Cornell University, Ithaca, NY, 14853, USA}
\affiliation{Universit\'e Paris Cit\'e, CNRS, Astroparticule et Cosmologie, F-75013 Paris, France}

% \maketitle
\begin{abstract}
    Extragalactic carbon monoxide (CO) line emission will likely be an important signal in current and future Cosmic Microwave Background (CMB) surveys on small scales. However, great uncertainty surrounds our current understanding of CO emission. We investigate the implications of this modeling uncertainty on CMB surveys. Using a range of star formation rate and luminosity relations, we generate a suite of CO simulations across cosmic time, together with the broadband cosmic infrared background (CIB). From these, we quantify the power spectrum signatures of CO that we would observe in a CMB experiment at 90, 150, and 220 GHz. We find that the resulting range of CO auto-spectra spans up to two orders of magnitude and that while CO on its own is unlikely to be detectable in current CMB experiments, its cross-correlation with the CIB will be a significant CMB foreground in future surveys. We then forecast the bias on CMB foregrounds that would result if CO were neglected in a CMB power spectrum analysis, finding shifts that can be comparable to some of the uncertainties on CMB foreground constraints from recent surveys, particularly for the thermal and kinetic Sunyaev-Zel'dovich effects and radio sources, and many times greater than the expected uncertainties expected for future data. Finally, we assess how the broad range of multifrequency CO$\times$CIB spectra we obtain is captured by a reduced parameter set by performing a principal component analysis, finding that three amplitude parameters suffice for a CMB-S4-like survey. Our results demonstrate the importance for future CMB experiments to account for a wide range of CO modeling, and that high-precision CMB experiments may help constrain extragalactic CO models.
\end{abstract}

\section{Introduction}
The Cosmic Microwave Background (CMB) has been a key workhorse of modern cosmology for the past several decades. Observations of the CMB, like those from \textit{Planck} \citep{collaboration_2020_Planck2018Results}, the Atacama Cosmology Telescope (ACT) (e.g., \citealt{choi_2020_AtacamaCosmologyTelescope,madhavacheril_2024_AtacamaCosmologyTelescope,ACT:2025fju}), and the South Pole Telescope (SPT) (e.g., \citealt{balkenhol_2023_MeasurementCMBTemperature}), alone have placed tight constraints on cosmological parameters, and are further tightened when combined with external datasets, like baryon acoustic oscillations (BAO) \citep[e.g.,][]{collaboration_2025_DESI2024III,collaboration_2025_DESI2024IV,alam_2017_ClusteringGalaxiesCompleted} and type Ia supernovae observations \citep[e.g.,][]{scolnic_2018_CompleteLightcurveSample}. The high resolution and sensitivity of future experiments, like Simons Observatory (SO) \citep{SimonsObservatory:2018koc} and CMB-S4 \citep{Abazajian:2019eic}, will be used to further extract cosmological information, much of which will come from measurements on scales of arcminutes. While the larger CMB scales are dominated by early universe astrophysics from $z\sim1100$, these arcminute scales are dominated by later-time processes. These low redshift signatures include the gravitational lensing of the CMB from the matter distribution in the universe; the thermal Sunyaev-Zel'dovich (tSZ) effect, which is the inverse Compton scattering of CMB photons with the hot gas from galaxies and galaxy clusters; and the kinetic Sunyaev-Zel'dovich (kSZ) effect, which is the effective Doppler shift of the CMB from scattering with the bulk flows of ionized gas \citep[see, e.g.,][for a review]{Bianchini:2025ksz}. The latter is of particular interest, as it is a rare probe of the epoch of reionization \citep[e.g.,][]{gruzinov_1998_SecondaryCMBAnisotropies,Knox:1998fp,Battaglia:2012im,Jain:2023jpy}. Emission from astrophysical sources is also significant in small-scale CMB data. At lower frequencies, radio sources, like active galactic nuclei, dominate. At higher frequencies, the total thermal emission from dusty, star-forming galaxies, or Cosmic Infrared Background (CIB), is the strongest signal in the extragalactic millimeter sky. These foregrounds, as well as Galactic dust emission, are currently considered and modeled in CMB power spectrum analyses \citep[e.g.,][]{dunkley_2013_AtacamaCosmologyTelescope,SPT:2020psp,balkenhol_2023_MeasurementCMBTemperature,Tristram:2023haj,ACT:2025fju,beringue_2025_AtacamaCosmologyTelescope}.

Another astrophysical foreground at these frequencies is the emission from extragalactic carbon monoxide (CO). Unlike the other foregrounds, CO does not have a continuum spectrum; rather, it emits at the specific frequencies that correspond to transitions between rotational states of the CO molecule. The transition CO($J\rightarrow J-1$) has a rest-frame frequency of $J \times 115.27$ GHz. The first two transitions emitted by nearby (i.e.\ very low redshift) CO therefore fall within common CMB bandpasses, the highest nominal central frequency of which is $\sim\,$220 GHz for recent ground-based surveys. This also means that higher CO transitions emitted from distant, large volumes in the universe are redshifted into the broad bandpasses of CMB surveys (see Fig.~\ref{fig:int_visualization}). While CO from the Milky Way has long been considered as a CMB foreground \citep{Planck:2013fzw}, extragalactic CO emission has not been as well explored. 

\begin{figure*}[t]
\includegraphics[width=\textwidth]{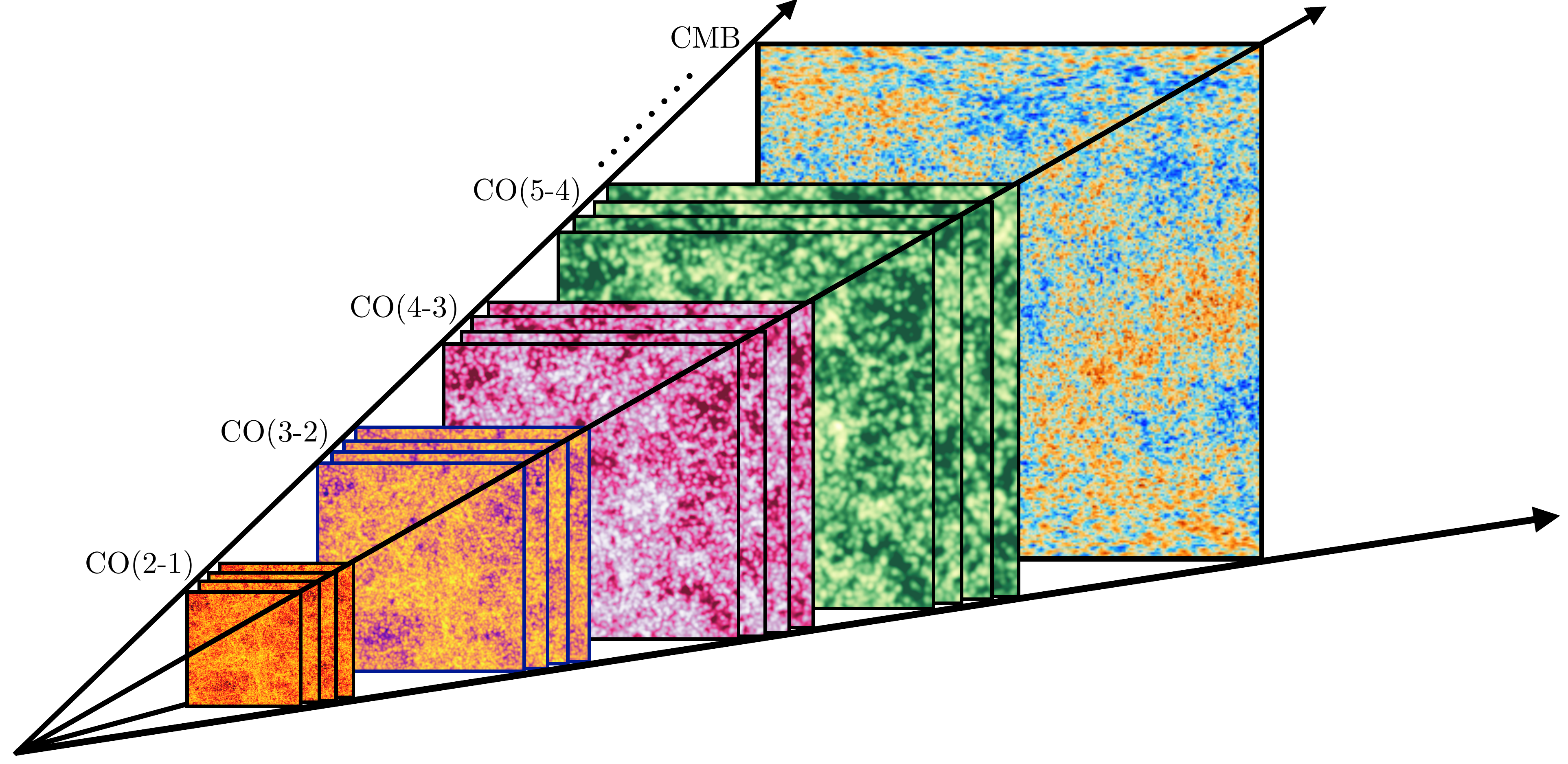} 
\caption{A lightcone visualization qualitatively depicting both the CO transitions in the foreground and CMB in the background. When observing the CMB, all CO transitions across cosmic time redshift into the bandpass and contribute to the total measured flux. While we do not use a lightcone approach, our power spectrum-based method is equivalent (see Sec.~\ref{sec:maps-and-power} for details).}
\label{fig:int_visualization}
\end{figure*}

The effect of extragalactic CO on CMB observations was first considered by \cite{Righi:2008br}, who argued that the autospectrum of these spectrally narrow line transitions would be suppressed as the bandpass of the instrument increased. They did not, however, consider cross-correlations between extragalactic CO and other CMB foregrounds. 

Recently, there has been renewed interest in the importance of extragalactic CO emission in CMB experiments. \cite{Maniyar:2023cuj}  and \cite{kokron_2024_ContributionsExtragalacticCO} (hereafter referred to as M23 and K24, respectively) added both CO and broadband far-infrared galaxy luminosities to N-body simulations of large-scale structure to obtain fully-correlated maps of CO and the CIB. This allowed them to compute the contributions of both CO and \cross power spectra to observations in CMB bands. While M23 used a single CO model, K24 considered a range of CO models based on observational uncertainties in the luminosity functions for each CO line transition from ALMA and NOEMA data. Both M23 and K24 found that the CO autospectrum would likely be undetectable in current ACT and SPT data. However, they also found that a much larger effect was the cross-correlation between extragalactic CO and the CIB, which they found to be a major CMB foreground after the CIB autospectrum. 

Current data are thus at the sensitivity and resolution that extragalactic CO may be a significant foreground. Additionally, \cite{Roy:2024kzc} recently measured CO(3$\rightarrow$2) at $z\sim0.5$ with a 3.2$\sigma$ significance by stacking \textit{Planck} CO maps, initially generated for studying the Milky Way, on BOSS galaxy positions. This marks the first potential measurement of extragalactic CO in CMB data. 

Astrophysical foregrounds and their cross-correlations, like the tSZ$\times$CIB cross-correlation \citep[e.g.,][]{Addison:2012my} contribute to the total measured power spectrum. To access individual signals, like the CMB or kSZ power spectra, these foregrounds must be accurately modeled and their parameters marginalized over. Neglecting or mismodeling any one of these foregrounds during the joint fit of all of the foreground and CMB power spectra may bias the results of CMB analyses, particularly on small scales. The kSZ auto-spectrum, for instance, was potentially detected statistically in SPT data with this method \citep{SPT:2020psp}, and tight limits were placed by ACT DR6 data \citep{ACT:2025fju}. These results were interpreted as constraints on the duration of reionization \citep{SPT:2020psp,beringue_2025_AtacamaCosmologyTelescope}. However, M23 and K24 found the \cross cross-correlation to be comparable to the kSZ signal. Since these kSZ constraints are already strongly correlated with the modeling of the tSZ~$\times$~CIB cross-correlation, it is possible that neglecting the contribution from other foregrounds, like CO~$\times$~CIB, to the total observed power may have biased these measurements. This is true of other foreground parameters as well: \cite{beringue_2025_AtacamaCosmologyTelescope} added templates from K24 of the CO and \cross spectra to their ACT DR6 analysis and found that some \cross models led to significant shifts in the best-fit values for foreground parameters, including kSZ and radio source parameters. Given that future CMB experiments will be even more sensitive \citep{SimonsObservatory:2018koc,Abazajian:2019eic}, a robust theoretical model of CO is required. 

More than a nuisance foreground, extragalactic CO is also a bright tracer of molecular gas and therefore star formation. As such, it is a major target for line intensity mapping (LIM) surveys, such as FYST \citep{CCAT-Prime:2021lly}, EXCLAIM \citep{pullen_2023_ExtragalacticScienceExperiment}, TIME \citep{crites_2014_TIMEPilotIntensityMapping}, and COMAP \citep{cleary_2022_COMAPEarlyScience}. Several models for CO luminosity have been developed. These encompass a diverse range of techniques, like semi-analytic and empirical modeling, and datasets, both simulated and observed \citep[see][for a review]{Bernal:2022jap}. This stems from the complexity of the connection between star formation and halo mass and the limited observations of individual CO line transitions, particularly at high redshift and at the faint end of the luminosity functions. The result is a nearly two-orders-of-magnitude uncertainty in the CO autospectrum \citep[e.g.,][]{Roy:2023cpx}. Therefore, a wide range of models must be considered to more robustly quantify the potential strength of CO in CMB surveys. While the CO power spectra of M23 and K24 are consistent with each other, the predictions from M23 are generally within the lower end of the range obtained by K24. As such, it is still unclear how representative the results of these analyses are of our current understanding of CO.

In this work, we use a wide range of CO models to predict the CO emission that would be seen in a CMB survey at the power spectrum level. The models correspond to various choices of both the star formation rate--halo mass relation and the infrared--CO luminosity relation. The models we consider were previously explored as signals for LIM surveys by \cite{Roy:2023cpx}. From these models, as well as a standard CIB model, we simulate the CO and CIB luminosity for each halo of an N-body simulation based on its mass and redshift. This expands upon the previous work of M23 and K24 by considering a much larger range of CO models, based on a variety of observational and simulated datasets. Given these simulations, we then convert the snapshots of our simulated box to a series of CO and CMB maps, from which we obtain the redshift evolution of the auto and cross spectra of each of our CIB and CO models. After integrating over redshift, we ultimately obtain CIB and CO power spectra and quantify the full effect that we expect extragalactic CO to have on a CMB survey within our models. 

Given this range of power spectra, we then turn to the broader implications for mm-wave surveys. We first use our resulting models to forecast the bias to CMB foreground measurements that we expect if \cross is neglected in a power spectrum analysis. We then assess whether the range of our models can effectively be spanned by a limited number of free parameters by performing a principal component analysis. 

The paper is structured as follows. In Section~\ref{sec: Modelling}, we detail the CO and CIB models that we consider in our analysis. In Section~\ref{sec: Sims}, we discuss the N-body simulations used and how we implement and measure the CO and CIB signals. In Section~\ref{sec:Results-Discussion}, we present and discuss the CO and \cross power and frequency spectra. In Section~\ref{sec:parameterbiases}, we perform a Fisher forecast to compute the foreground parameter biases.  In Section~\ref{sec:PCA}, we obtain the PCA-based templates to capture the range of anticipated \cross spectra. We conclude in Section~\ref{sec: Conclusion}.

\section{Theoretical Modeling} \label{sec: Modelling}

\subsection{Extragalactic Carbon Monoxide} \label{subsec: CO models}

Carbon monoxide (CO) rotational lines are an important tool for studying the molecular gas in galaxies, which is essential for understanding star formation and, therefore, galaxy formation and evolution. Several models have been developed to predict the power spectrum expected from CO, especially in anticipation of upcoming LIM experiments. The models we consider are implemented in the publicly available code \verb|LIMpy| \citep{Roy:2023cpx} and are based on various observational and simulated datasets. We describe these models below.

We begin with the star formation rate (SFR) of a halo. The relationship between halo mass and SFR is complex and poorly constrained, particularly for low halo masses and luminosities. Galaxy chemical composition, morphology, mergers, environment, and feedback all contribute to the SFR over cosmic time \citep[see][for a review]{blanton_2009_PhysicalPropertiesEnvironments,jr_2012_StarFormationMilky}. To capture a broad range of physical processes, we consider five different SFR models. \verb|Silva15| \citep{silva_2015_PROSPECTSDETECTINGII} and \verb|Fonseca16| \citep{fonseca_2017_CosmologyIntensityMapping} both model the \masshalo -- SFR relation as a power law with parameters fit to a simulated galaxy catalog with a minimum halo mass of $10^8 \, M_{\odot} / h$. This catalog was created by applying a semi-analytic stellar population synthesis model with dust attenuation 
\citep{delucia_2006_FormationHistoryElliptical, delucia_2007_HierarchicalFormationBrightest, guo_2011_DwarfSpheroidalsCD} to the dark matter-only Millenium \citep{springel_2005_SimulationsFormationEvolution} and Millennium II \citep{boylan-kolchin_2009_ResolvingCosmicStructure} N-body simulations. The two models differ in the number of terms and parameters used for their \masshalo -- SFR relations. As a result, \verb|Silva15| is applicable for redshifts of $z < 20$, while \verb|Fonseca16| is modeled for $z < 10$ and then held fixed for $10 < z < 20$. The next models we consider are \verb|TNG100| and \verb|TNG300| \citep{TNGa,TNGb,TNGc,TNGd,TNGe}, which capture the mean SFRs obtained from the hydrodynamical IllustrisTNG simulations of the same names. The primary difference between these simulations are their volumes (and therefore resolutions), which are $100^3$ and $300^3$ Mpc$^3$ each. Finally, the \verb|Behroozi19| \citep{behroozi_2019_UniverseMachineCorrelationGalaxy} model uses the \verb|UniverseMachine| simulations, which fit SFR to halo properties based on observations, like the galaxy UV luminosity functions, observed stellar mass functions, and different galaxy populations, to obtain the SFR for $z<10$. Each of these \masshalo -- SFR relations are shown in Fig.~3 of \cite{Roy:2023cpx}.

From these SFRs, we obtain the luminosity of CO lines for a given dark matter halo as
\begin{equation} \label{eq: L_co}
\log_{10}[L_{\text{CO}(J)}] = a_{\text{CO}(J)} + b_{\text{CO}(J)}  \log_{10}[\text{SFR}(M_{\mathrm{halo}}) ],
\end{equation}
where $L_{\text{CO}(J)}$ represents the luminosity of the CO transition from rotational level $J$, SFR(\masshalo) is the SFR as a function of halo mass, and $a_{\text{CO}(J)}$ and $b_{\text{CO}(J)}$ are constants that reflect the empirical calibration of the line emission. Both the SFR(\masshalo) and coefficients are modeled independently. 

We obtain these coefficients by applying three different SFR -- $L_\mathrm{co}$ scaling relations for a given SFR. Each are based on different space- and ground-based datasets. The \verb|Greve14| \citep{greve_2014_STARFORMATIONRELATIONS} model uses galaxy observations from \textit{Herschel}/SPIRE, the Infrared Astronomical Telescope (IRAS), IRAM 30-m Telescope, and the James Clerk Maxwell Telescope (JCMT) \citep{Papadopoulos:2011kb}. These galaxies have infrared luminosities of $\ge10^{11} \; L_\odot$ and were observed up to $z\le6.3$. \verb|Kamenetzky15| \citep{kamenetzky_2016_RELATIONSCOROTATIONAL} also uses galaxy spectra from \textit{Herschel} for the higher CO transitions and those from the Arizona Radio Observatory (ARO) for the lower transitions. These observations are at $z<1$. Finally, the \verb|Visbal10| \citep{Visbal:2010rz} model is calibrated to spectral measurements of M82 taken with \textit{Herschel} \citep{Panuzzo:2010qc}. This diverse range of models will allow us to quantify the full range of possible CO signals.

\subsection{Cosmic Infrared Background} \label{subsec: CIB model}

\begin{table*} 
    \centering
	\begin{tabular}{|c|c|c|}
		\hline
		S12 Parameter & Description & Value\\
		\hhline{|=|=|=|}
		$L_o$ & Overall luminosity normalization & $1.59\times10^{-15} \;L_\odot/M_\odot$ \\
		\hline
		$M_{\mathrm{eff}}$ & Most luminous halo mass & 12.6 $M_\odot$ \\
		\hline
		$\delta$ & Redshift evolution power law index & 3.6 \\
		\hline
		$\beta$ & Low frequency emissivity index & 1.75 \\
		\hline
		$\gamma$ & High frequency SED power law index & 1.7 \\
		\hline
		$T_o$ & Dust temperature at $z=0$ & 24.4 K  \\
        \hline
		$\alpha$ & Dust temperature power law index & 0.36 \\
		\hline
	\end{tabular}
    
    \caption{The parameter values used for the S12 CIB model \citep{Shang:2011mh}. Values were obtained from fitting to \textit{Planck} and IRAS CIB autospectrum data \citep{Planck:2013wqd,McCarthy:2020qjf}.}
    
    \label{tab:cib_params}
    
\end{table*}

Another tracer of the star formation history is the CIB. We implement the CIB model from \cite{Shang:2011mh}, hereafter known as S12. While somewhat simplistic, this model yields predictions that agree with the current most precise measurements of the CIB power spectrum, which come from \textit{Planck} \citep{Planck:2013wqd} and \textit{Herschel} \citep{Viero:2012uq}. Consequently, it has been widely used in various analyses over the past decade \citep{Zahn:2011vp,tucci_2016_CosmicInfraredBackground,stein_2020_WebskyExtragalacticCMB,lembo_2022_CMBLensingReconstruction, Rotti:2022lvy} and continues to be utilized by current CMB analyses \citep[e.g.,][]{Wenzl:2024sug, ACT:2023oei, Kim:2024dmg,maccrann_2023_AtacamaCosmologyTelescope,Mondino:2024rif}. Given its use in modern cosmological analyses, we include it in this study. 

In this model, we ascribe an infrared luminosity to each halo based on its mass and redshift as well as the observing bandpass.  The far-infrared luminosity for a given halo is factored into its dependencies on redshift, halo mass, and frequency according to
\begin{equation}
    L^{\mathrm{IR}}_{\nu_{\mathrm{rest}}} = L_o  \Phi(z) \Sigma(\masshalo) \Theta(\nu_{\mathrm{rest}})   ,
\end{equation}
where $L_o$ is an overall normalization, $\Sigma$ is the luminosity-halo mass ($L-$\masshalo) relation, $\Phi$ is the redshift dependence for the $L-\masshalo$ relation, and $\Theta$ is the rest frame SED of the galaxy. The rest frame frequency is related to the observed frequency $\nu$ according to $\nu_{\mathrm{rest}} = \nu (1+z)$. 

The $L-\masshalo$ relation,
\begin{align*}
    \Sigma(\masshalo) = & \frac{\masshalo}{\sqrt{2\pi \sigma^2_{L/M}}} \numberthis  \\
    & \times \exp\left( - \frac{(\log_{10} \masshalo - \log_{10}M_{\mathrm{eff}})^2} {2\sigma_{L/M}^2} \right) ,
\end{align*}
is a mass-weighed lognormal function parameterized by two parameters: the mean $M_\mathrm{eff}$ and variance $\sigma^2_{L/M}$ of the distribution. This suppresses star formation at both the lower mass range and the high mass range. The physical mechanisms for the latter include photoionization heating, supernovae, active galactic nuclei, and virial shocks. 

The redshift evolution of this $L-\masshalo$ relation is given as
\begin{equation}
	\Phi(z) = (1+z)^{\delta}
\end{equation}
for a given power $\delta$. Since the halo-to-stellar mass ratio evolves only mildly with redshift \citep{Neistein:2011hk} and the star formation rate (SFR) for dusty, star forming galaxies is related to their infrared luminosity \citep{Kennicutt:1998zb,Carilli:2011vw}, the redshift evolution of the $L-\masshalo$ relation can be traced by that of the specific star-formation rate ($\mathrm{SFR}/M_{\mathrm{*}}$). While this scaling is well-motivated by semi-analytic models, observations tend to measure a plateau for $z \gtrsim 2$ (see \citealt{Weinmann:2011dq} for a compilation); however, some observations measure an SFR that continues to rise earlier than $z\sim2$ \citep{Yabe:2008gg}. For this work, we follow S12 and do not adopt a plateau for simplicity.

Finally, the following spectral energy density (SED) is assumed for all halos and subhalos:
\begin{equation}\label{eqn:cib_sed}
    \Theta (\nu,z) = \begin{cases}
        \left(\frac{\nu}{\nu_o} \right)^{\beta} \left(\frac{B_{\nu}(T_{\mathrm{dust}})}{B_{\nu_o}(T_{\mathrm{dust}})} \right), & \nu < \nu_o; \\
        \left(\frac{\nu}{\nu_o} \right)^{-\gamma}, & \nu > \nu_o,
    \end{cases}
\end{equation}
where $B_\nu$ is the Planck function. At low frequencies, the SED is a modified blackbody, which accounts for the effect of dust on the SED. At high frequencies, a power law function is adopted to more rapidly temper the Wien tail. The rest frame frequency at which these two SED sections smoothly connect, $\nu_o$, satisfies the condition
\begin{equation}
	\left. \frac{d\ln\Theta(\nu_{\mathrm{rest}}, z)}{d\ln\nu_{\mathrm{rest}}} \right|_{\nu_{\mathrm{rest}} = \nu_o} = -\gamma.
\end{equation}
The dust temperature $T_\mathrm{dust}$ evolves in redshift according to
\begin{equation}
    T_d = T_o (1+z)^{\alpha}
\end{equation}
for parameters $T_o$, which represents the dust temperature in these galaxies at $z=0$, and $\alpha$.

We choose the parameter values from \cite{Planck:2013wqd} and \cite{McCarthy:2020qjf} (see Table~\ref{tab:cib_params} for values). In their analysis, \cite{Planck:2013wqd} held $\sigma^2_{L/M}$ fixed at 0.5 and added an additional parameter, $M_\mathrm{min}$, which is the minimum halo mass that can produce CIB luminosity. This parameter is between $10^{10} - 10^{11} \;M_\odot$ and was unconstrained by \cite{Planck:2013wqd}. We do not impose such a minimum halo mass in our CIB model and are constrained only by the minimum halo mass of our simulations (see Sec.~\ref{subsec: n-body}). They then fit these 8 parameters and shot noise amplitudes for each of their multifrequency CIB power spectra (an additional 15 parameters) from \textit{Planck} and IRAS \citep{Miville-Deschenes:2004tjd} CIB data.

\section{Simulations to Power Spectra}  \label{sec: Sims}

We simulate maps of CO and CIB over a wide span of cosmic time. To obtain this, we apply our CO and CIB models to 3D snapshots of N-body simulations at various redshifts and sum along a chosen line-of-sight within each snapshot to generate 2D intensity maps of the CIB and each CO line. We then compute the power spectra of the CO, CIB, and their cross-correlation at each snapshot before finally integrating across redshift to obtain the contribution of CO and the CIB to the total power spectrum of a CMB survey. We detail this process below.

\subsection{Adding CO and CIB Emission to N-body Simulations} \label{subsec: n-body}

To generate simulations of CO and CIB emissions, we first utilize the \texttt{MP-GADGET} software package \citep{Springel:2020plp} to perform $N$-body simulations of dark matter. We save the simulations at 100 snapshots between $0 \le z \le 20$, each representing an equal time interval of approximately 130 Myr (corresponding to the light travel time across the simulation volume). We simulate the dark matter with a minimum particle mass of $6.43$\sci{8} $M_\odot/h$ in a cubic box with side length $100$ comoving Mpc (cMpc) per $h$. We set the initial conditions at $z \sim 200$ and evolve to lower redshifts of interest. The cosmological parameters we selected for these simulations are consistent with the results from the Planck collaboration \citep{Planck:2015fie}.

All 100 redshift slices correspond to a single ``simulation set". We employ 13 such sets of our simulated box, each with 100 snapshots (for a total of 1,300 snapshots across our entire simulation suite) and initial conditions generated from a distinct random seed. These different realizations of initial conditions simulate sample variance. For each snapshot, we generate halo catalogs with a minimum halo mass of $2.05$\sci{10} $M_\odot/h$ using the friends-of-friends (FoF) algorithm with a linking length of 0.2. These halo catalogs form the basis for creating intensity maps, which we generate using the \texttt{LIMpy} package \citep{Roy:2023cpx}. This tool transforms the halo catalogs into intensity maps based on user-defined parameters, facilitating the simulation of line intensity mapping (LIM) and CIB signals. While the line modeling that we employ in this analysis has already been implemented in \texttt{LIMpy} and used in previous works, we implemented CIB modeling into \texttt{LIMpy} for the first time for this project. 

For each snapshot, we use \verb|LIMpy| to assign a CO and CIB luminosity to each halo in our simulation based on the models from Section~\ref{sec: Modelling}. We then form a regular 3D grid of luminosities by summing the luminosity within each voxel. We convert from luminosity to intensity using
\begin{equation}
    I(\vec{\theta},\chi) =  \frac{L(\vec{\theta},\chi)}{4 \pi D_{L}^{2}(z)\Delta\Omega\Delta\nu_{\mathrm{bandpass}}} ,
\end{equation}
where $\vec{\theta}$ and $\chi$ are the angular position and comoving distance along the line of sight from the observer to each voxel, $I$ and $L$ are the intensities and luminosities of each voxel, $D_L$ is the luminosity distance to the redshift of that snapshot, $\Delta\Omega$ is the angular size of the voxel, and $\Delta\nu_{\mathrm{bandpass}}$ is the spectral width of the observed bandpass. The units are set by the conversion factor 
\begin{equation}
    4.0204\times 10^{-2} h^2 \left[ \frac{\mathrm{Jy}}{\mathrm{sr}} \middle/ \frac{L_{\odot}}{(\mathrm{Mpc} \cdot h^{-1})^{2} \;\mathrm{sr} \cdot \mathrm{GHz} } \right].
\end{equation}
With this, we obtain the 3D intensity grids in Jy/sr for each CO and CIB model.

\subsection{Maps and Power Spectra of CO and CIB} \label{sec:maps-and-power}

There are two common techniques to approximate the CO and CIB power spectra from the simulations. The first, which was taken by M23 and K24, is to construct a lightcone, integrate the simulations over redshift within the lightcone to obtain a single map, and calculate the power spectrum of that map. The other method is to calculate the contribution to the power spectra from each snapshot and combine the spectra to obtain the total expected power spectrum \citep[e.g.,][ used this method for calculating SZ power spectra from simulations]{battaglia_2012_ClusterPhysicsSunyaevZeldovich}. The latter method is less memory-intensive and requires taking Fourier transforms across smaller volumes than a lightcone, thereby significantly speeding up our analysis. This ``snapshot-level" method also allows us to determine the power spectrum from large regions of the universe out to high redshift using our relatively modest box size with high mass resolution. For these reasons, we employ this ``snapshot-level" approach.

We choose a specific axis of each intensity grid, or snapshot, to act as the line of sight and sum the intensities along that axis to create a 2D intensity map at that redshift. Since we take each grid to correspond to a single redshift, these 2D maps approximate line-of-sight integrals of narrow slices of the lightcone centered on the redshifts of each box. At the lowest redshift of $z=0.01$, our map size is 191.4 deg$^2$, and at the highest redshift of $z=11.98$, the map size is 0.8 deg$^2$. This allows us to consider a multipole range of $200 \le \ell \le 12000$. Our simulations are thus large enough to capture the Poisson regime and some of the clustering regime of the power spectra of interest. We implement flux cuts of 15 mJy to each map, corresponding to an ACT-like CMB survey \citep[e.g.,][]{ACT:2025fju} (though we note that SPT analyses in \citealt{SPT:2020psp} and \citealt{balkenhol_2023_MeasurementCMBTemperature} use different thresholds). Each pixel in the map in excess of the flux threshold is replaced by the average flux in the map. Examples of these simulated maps for several CO and CIB models are shown in Figure~\ref{fig:sims}.

%Sims
\begin{figure*}
    \includegraphics[width= \textwidth]{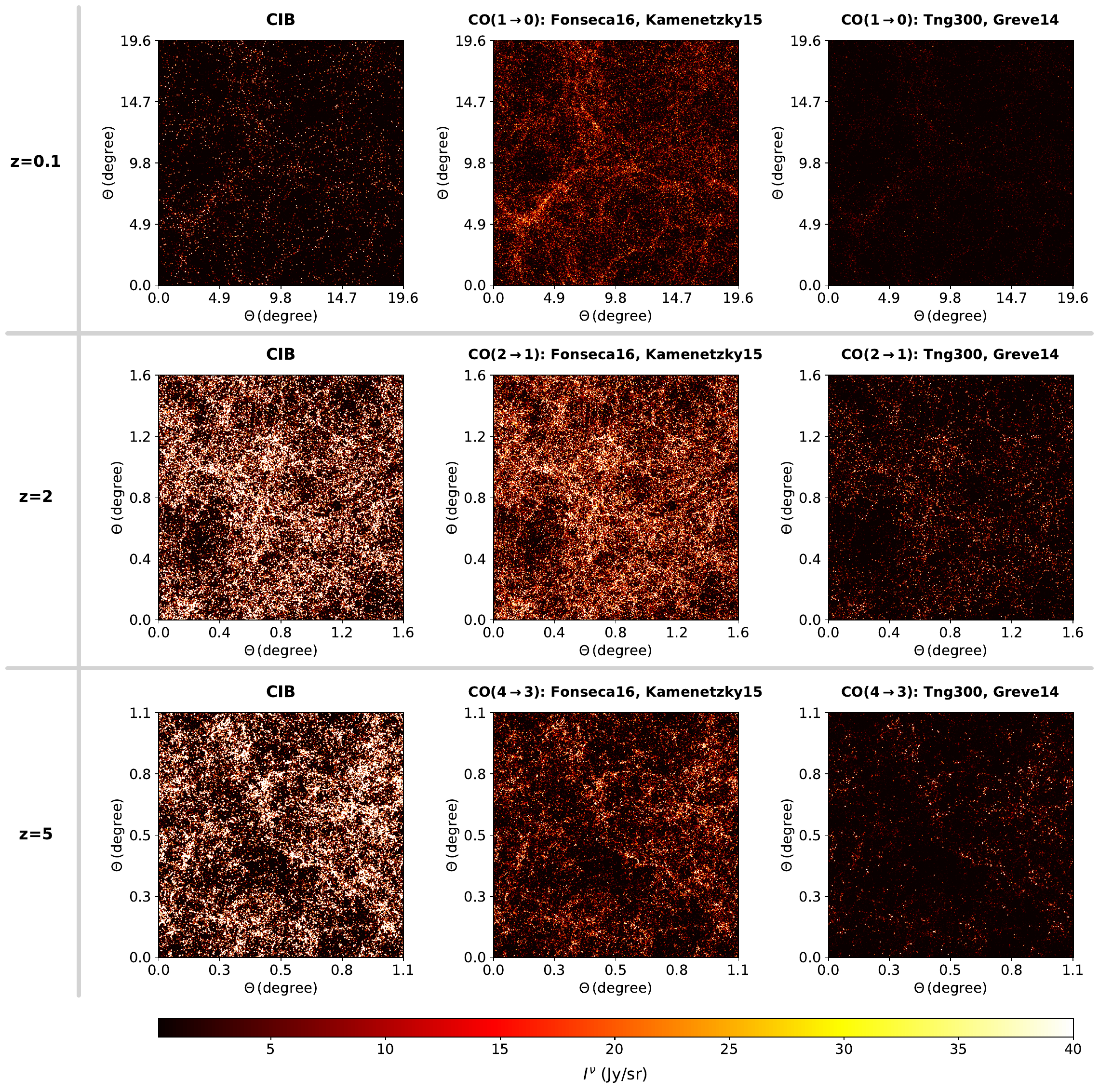}     
    \caption{Our simulated maps of the S12 CIB model and two CO models for several redshifts that contribute to our considered 90 GHz bandpass. For a given CO line, the bandpass selects the source redshifts, while for the CIB, all redshifts contribute to the observed emission. The CO intensity varies significantly with the model choice and is subdominant to the CIB at each snapshot. Both the CIB and CO intensity peak at cosmic noon around $z \sim 2$.} 
    \label{fig:sims}
\end{figure*}

The contribution to the observed power spectrum from a given source (e.g.,\ CO$\times$CO or CO$\times$CIB) is the sum of all contributions that redshift into the chosen bandpass; these contributions will arise from different line-of-sight distances from the observer, and their sum can therefore be written as
\begin{equation}
C_\ell^{\rm obs} = \int \frac{dC_\ell^{\rm obs}}{d\chi} d\chi\ .
\label{eq:Cellobs-integral}
\end{equation}

Each one of our 2D intensity maps is itself the sum of contributions to the intensity from a range of line-of-sight distances, so we can approximate $dC_\ell^{\rm obs}/d\chi$ at the redshift of a simulation snapshot by measuring the angular power spectrum of the map (or cross power spectrum between two maps) at that redshift $C_\ell^s$ and dividing by the comoving line-of-sight width of the box:
\begin{equation}
\frac{dC_\ell^{\rm obs}}{d\chi}(z_s) \approx \frac{C_\ell^s}{\Delta\chi_{\rm box}} ,
\end{equation}
where $\Delta\chi_{\mathrm{box}}$ is the length of the comoving box along the line-of-sight (which is 100 cMpc$/h$ for our simulations) and index $s$ denotes each snapshot.
We use the flat-sky approximation to calculate the angular power spectrum of a given snapshot as follows:
\begin{equation}
    \langle I^s(\vec{\ell}_{1}) I^s(\vec{\ell}_{2}) \rangle = (2 \pi)^{2} \delta(\vec{\ell}_{1}+\vec{\ell}_{2}) C^s_{\ell_{1}},
\end{equation}
where $I^s(\vec{\ell})$ is the Fourier transform of the intensity of the map. The integral in Eq.~\eqref{eq:Cellobs-integral} can then be approximated by the following Riemann sum:
\begin{equation}
\int \frac{dC_\ell^{\rm obs}}{d\chi} d\chi
	\approx \sum_s \frac{C_\ell^{s}}{\Delta\chi_{{\rm box}}} 
	(\bar{\chi}_{{\rm box},\;s} - \bar{\chi}_{{\rm box},\;s-1})\ ,
\label{eq:Cellobs-integral-Riemann}
\end{equation}
where $\bar{\chi}_{{\rm box,}\;s}$ is the comoving distance from the observer at $\chi=0$ to the redshift of the snapshot.\footnote{The integral can also be evaluated using another approximate scheme, such as the trapezoid rule, but we find that the results are equivalent to the Riemann sum in Eq.~\eqref{eq:Cellobs-integral-Riemann} at the precision relevant for our investigation.} In this computation, each adjacent snapshot belongs to a different simulation realization, and we only repeat realizations once every 14 snapshots; this mimics a lightcone with distinct structures at different line-of-sight distances.

Since the CIB is comprised of spectrally continuous sources, all redshifts contribute to each frequency band and so contribute to the summation in Eq.~\eqref{eq:Cellobs-integral-Riemann}. To account for the total intensities coming from the various CO line emissions to a particular frequency band, we consider all CO lines that fall within the same observed frequency channel for a given experimental bandwidth $\Delta \nu$. These lines are sourced at various redshifts. The observed frequencies $\nu_\mathrm{obs}$ are related to the source frequency emitted in the rest frame as 
\begin{equation}
    \nu_{\mathrm{obs}} = \frac{\nu_\mathrm{rest}}{1+z}.
\end{equation}
For example, an experiment observing at 90 GHz with a 60 GHz bandwidth would detect CO(2-1) emissions from redshifts~$0.92$ to~$2.83$ but would detect CO(3-2) emissions from redshifts~$1.87$ to~$4.75$. We assume a tophat bandpass of 50 GHz centered on 3 common CMB observing frequency channels: 90, 150, and 220 GHz. These bandpasses were also used in M23, making our results directly comparable. K24 used the exact ACT bandpasses, which are most similar to our assumed bandpass at 150 GHz. Notably, multiple CO transitions emitted at the same redshift may contribute to the power spectrum of a single bandpass. For each CO line, we determine the snapshots that contribute to a given bandpass before generating maps for those CO lines individually at those redshifts. The redshift ranges for each line are shown in Fig.~\ref{fig:dcdz}. These snapshots are the ones used in the summation in Eq.~\eqref{eq:Cellobs-integral-Riemann}

We compute all combinations of power spectra that contribute to the 2-point function of the maps. Within a single map at a given redshift that contributes to a frequency $\nu$, the total intensity at a given angular position in the map $I^{s_\nu}(\vec{\theta})$ is
\begin{equation} \label{eq:intensity_map}
    I^{s,\nu}(\vec{\theta}) = I^{\mathrm{CIB}_{s,\nu}}(\vec{\theta}) + \sum_{j} I^{\mathrm{CO}(j+1 \to j)_{s,\nu}}(\vec{\theta}) ,
\end{equation}
where $I^{s,\,\mathrm{CO}(j+1 \to j)_\nu}$ is the intensity of line CO($j+1\to j)$ that contributes to bandpass $\nu$. The total power spectrum of the map at a single frequency from the combination of CO and the CIB is then
\begin{align*}
    C_{\ell}^{s,\nu} 
    	=&\; C_{\ell}^{\mathrm{CIB}_{s,\nu}} \\
    &\!+\sum_{j} \left[ 
    	C_{\ell}^{\mathrm{CO}(j+1 \to j)_{s,\nu}} 
	+ 2 C_{\ell}^{\mathrm{CO}(j+1 \to j)_{s,\nu}\times \mathrm{CIB}_{s,\nu}} 
    \right] \\ 
    & \!+\sum_{j}\sum_{k, \,j \neq k} 
    	C_{\ell}^{\mathrm{CO}(j+1 \to j)_{s,\nu} \times \mathrm{CO}(k+1 \to k)_{s,\nu}}\ .
    \numberthis
    \label{eq:ps_single_nu}
\end{align*}
Here, the first term is the CIB autospectrum; the second is the sum of the autospectra of each CO line and the cross-correlations between each CO line and the CIB. The third is the sum of the cross-correlations between the CO lines present in the map. We also compute the cross-frequency power spectra of the maps by generalizing Eq.~\eqref{eq:ps_single_nu} as
\begin{widetext}
\begin{equation}\label{eq:ps_multi_nu}
\begin{split}
    C_{\ell}^{s,\nu1 \times s, \nu2} = &\; C_{\ell}^{\mathrm{CIB}_{s,\nu1}\times\mathrm{CIB}_{s,\nu2}} \\
    & +\sum_{j} \left[ C_{\ell}^{\mathrm{CO}(j+1 \to j)_{s,\nu1}\times\mathrm{CO}(j+1 \to j)_{s,\nu2}} + C_{\ell}^{\mathrm{CO}(j+1 \to j)_{s,\nu1}\times \mathrm{CIB}_{s,\nu2}} + C_{\ell}^{\mathrm{CO}(j+1 \to j)_{s,\nu2}\times \mathrm{CIB}_{s,\nu1}} \right] \\ & +\sum_{j}\sum_{k, \,j \neq k} C_{\ell}^{\mathrm{CO}(j+1 \to j)_{s,\nu1} \times \mathrm{CO}(k+1 \to k)_{s,\nu2}}\ .
\end{split}
\end{equation}
\end{widetext}
We compute these power spectra for each CO model. Throughout this work, we present all power spectra in CMB temperature units of $\mu {\rm K}^2$.

\section{Simulation Results} \label{sec:Results-Discussion}

\subsection{Redshift Evolution of CO}
%dC/dz
\begin{figure*}
    \centering
    \includegraphics[width= 0.98\textwidth]{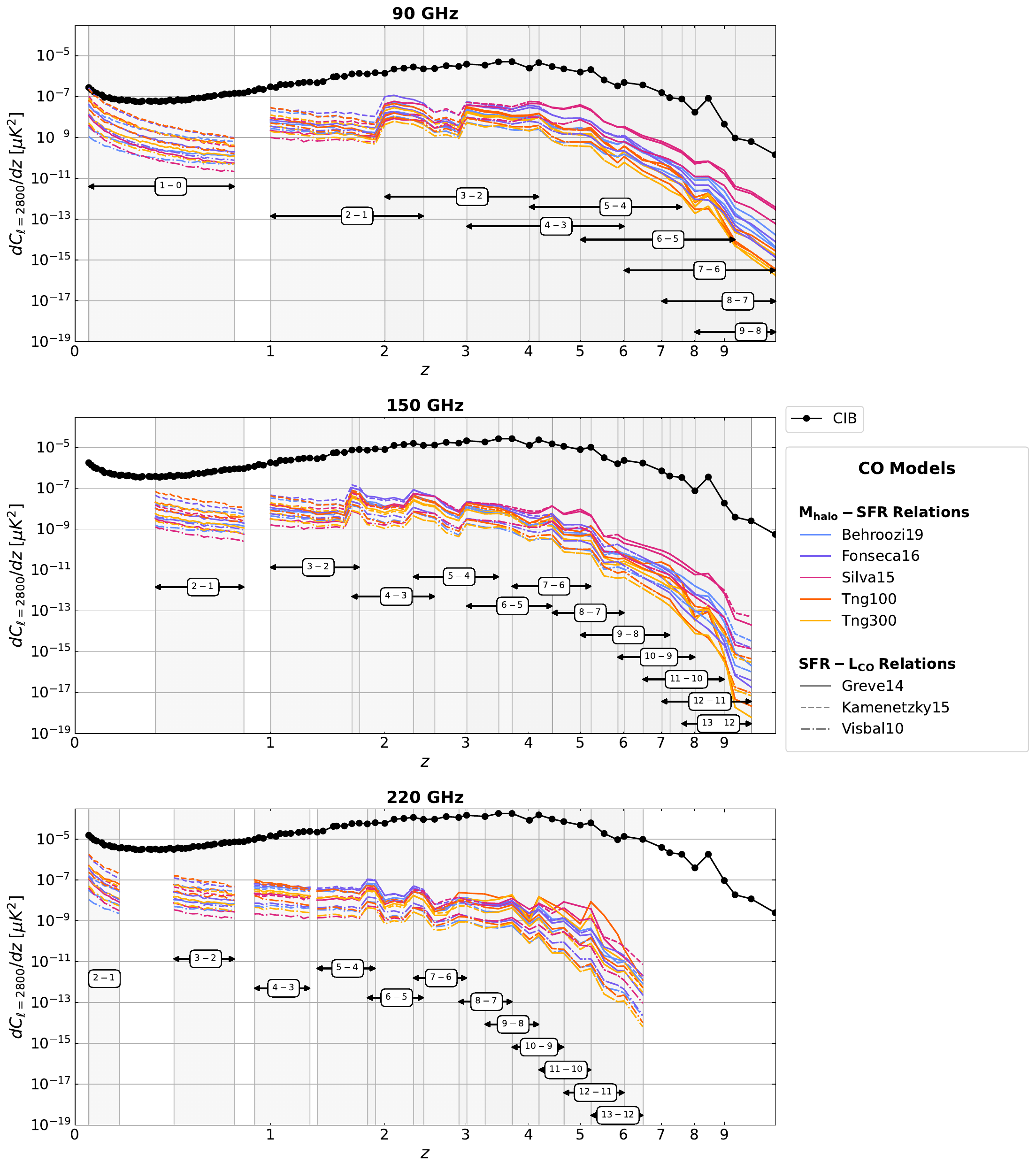}
    \caption{Auto-spectra of CO and CIB from our simulations observed at 90, 150, and 220 GHz and $\ell=3000$ as functions redshift. The grey bands correspond to the redshifts that source the various CO lines that contribute to a given observing frequency; the transitions are noted in text boxes centered on their respective grey band. The width of the grey bands corresponds to that of the broad CMB bandpasses of $\Delta \nu = 50\,{\rm GHz}$, resulting in contributions to CO from almost all considered redshifts at the lower observing frequencies. The similar redshift evolution between the CO and CIB, particularly with their peak at cosmic noon, results from the fact that they are both tracing the star formation history of the universe. As a result, their cross-correlation is appreciable.}
    \label{fig:dcdz}
\end{figure*}

By computing the angular power spectra of our simulations at each snapshot, we obtain the redshift evolution of the CO auto-spectra, $dC_\ell/dz$, shown in Fig.~\ref{fig:dcdz}. We plot the CO auto-spectra from each line individually and explicitly show the redshift range that sources each transition, which then redshifts into the observed bandpass. The broad CMB bandpasses pick up large volumes of the universe for each CO line transition, resulting in contributions from a wide range of redshifts. Only the 90 GHz bandpass is sensitive to all redshifts in the nearby universe ($z\lesssim1$), whereas the higher frequencies have gaps in their low-redshift contributions. These broad redshift ranges also overlap so that the same sources emit multiple CO lines. Most redshifts above $z\sim1.5$ source multiple CO lines for each considered frequency, particularly at higher redshifts. Because of this near-continuous redshift coverage for all but the lowest redshifts, the redshift evolution of CO in CMB surveys is more similar to that of a continuum source than a line emission. At each redshift, the contribution to the observed CO power spectrum varies considerably from the different modeling choices.

We also show the redshift evolution of the CIB auto-spectrum in Fig.~\ref{fig:dcdz}. The redshift evolution between the CIB and CO is similar (peaking at cosmic noon, around $z\sim2$) because they both trace the history of star formation. Note that while CO is sourced at higher redshifts, redshifts beyond $z\sim3-4$ are several orders of magnitude smaller than the contributions from cosmic noon and local redshifts. This redshift evolution similarity suggests a considerable cross-correlation between the two astrophysical signals.

\subsection{Power Spectra}

%Dl's
\begin{figure*}
    \centering
    \includegraphics[height= 0.88\textheight]{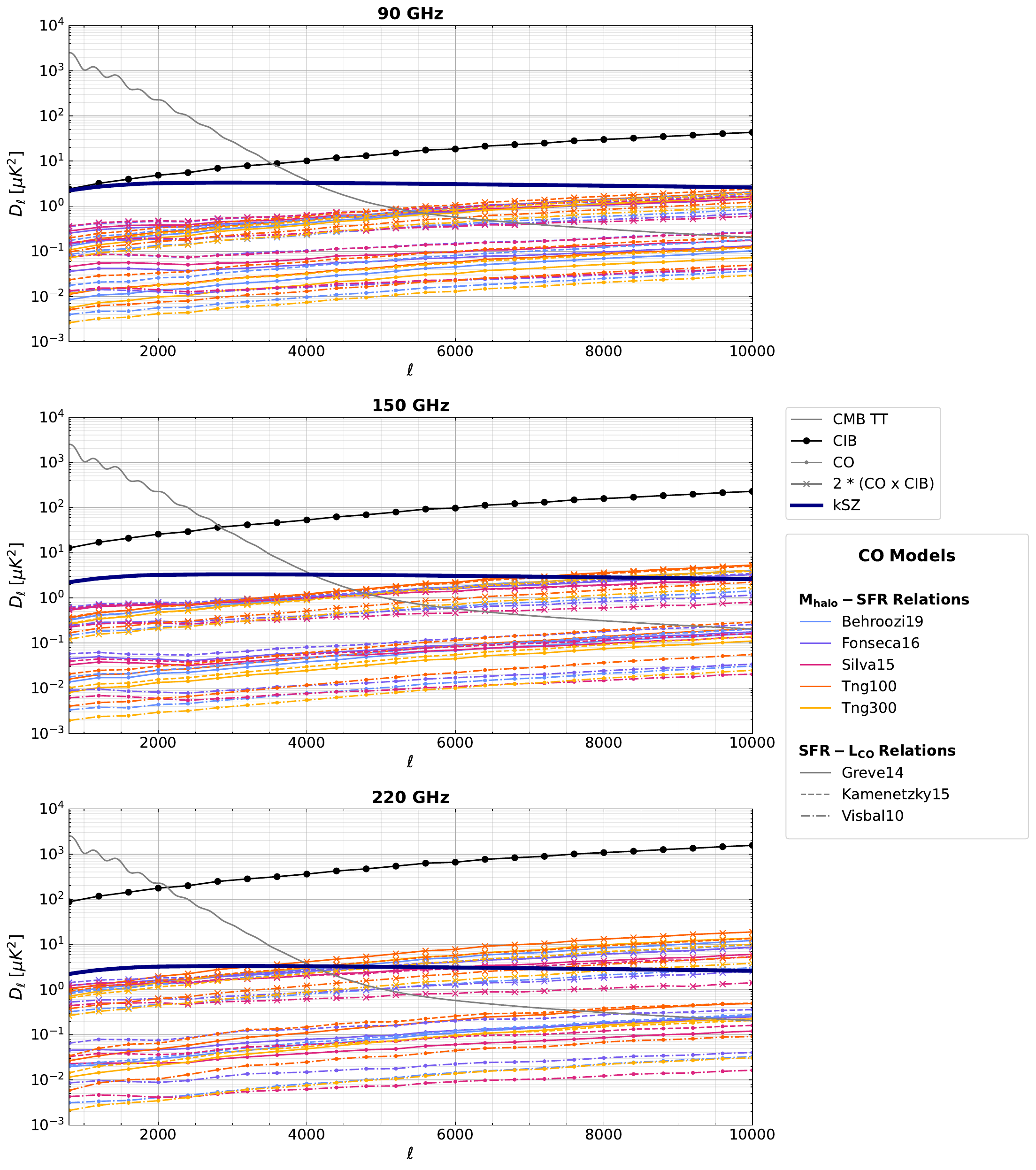}
    \caption{Auto-power spectra observed at 90, 150, and 220 GHz of CO and CIB and their cross-spectra from our simulations for the variety of models considered, shown as $\mathcal{D}_\ell \equiv \ell(\ell + 1)C_\ell/(2\pi)$. Their cross-correlation includes both cross terms, shown in Eq.~\eqref{eq:ps_single_nu}. The range of CO auto-spectra spans about 1.5 to 2 orders of magnitude at all frequencies and $\ell$'s, and their cross-spectra with the CIB span about 1 order of magnitude. The CO auto-spectra and cross-spectra with the CIB are comparable on small scales to the lensed primary CMB and the kSZ power spectrum (also shown here), respectively.} 
    \label{fig:dls}
\end{figure*}

Integrating over redshift, we obtain the total expected power spectra for CO, the CIB, and their cross-correlation (see Fig.~\ref{fig:dls}) at our observed frequencies. Our range of CO auto-spectra spans $\sim 1 - 2$ orders of magnitude. This range is relatively consistent across all multipoles. The \cross range consistently spans about one order of magnitude at all multipoles. For all multipoles, the \verb|Visbal10| models regularly predict lower power for both the CO auto-spectrum and \cross, while the \verb|Kamenetzky15| and \verb|Greve14| models are comparable to each other.

We also plot the expected kSZ auto-spectrum, which was obtained from the publicly available code \verb|fgspectra|\footnote{\url{https://github.com/simonsobs/fgspectra/tree/main}}. This template includes contributions from both late times and reionization. The former was derived from hydrodynamical simulations that included radiative cooling, star formation, and feedback from supernovae and AGN \citep{battaglia_2012_ClusterPhysicsSunyaevZeldovich}; the latter was derived from a semi-analytic model of the epoch of reionization calibrated to radiation-hydrodynamical simulations \citep{Battaglia:2012im}. This template has been used to obtain observational constraints on kSZ \citep{dunkley_2013_AtacamaCosmologyTelescope}.

\begin{table}
    \centering
    \begin{tabular}{c||c|c|c}
    	\multicolumn{4}{c}{\textbf{CO Autospectrum Amplitudes}} \\
	\hline \hline
         Frequency & M23 & K24 & This Work \\
         (GHz) & ($\mu$K$^2$) & ($\mu$K$^2$) & ($\mu$K$^2$) \\
         \hhline{= =|=|=}
         90  & --- & 0.05 -- 0.28  & 0.0054 -- 0.083 \\
         150  & 0.08  & 0.02 -- 0.2  & 0.0037 -- 0.061  \\
         220  & 0.5  & 0.08 -- 0.36  & 0.0049 -- 0.105  \\
    \end{tabular}
    \caption{Amplitudes $\mathcal{D}_\ell$ of the CO auto-spectra at $\ell=3000$ for common CMB observing frequencies. We compare the results of this work to those of M23 and K24. Our range of CO auto-spectra are generally lower than those from previous works.}
    \label{tab:co_auto}
\end{table}

\begin{table}
    \centering
%    \caption*{CO x CIB Amplitudes}
    \begin{tabular}{>{\hspace{2pt}}c||c|c|c<{\hspace{2pt}}}
    	\multicolumn{4}{c}{\textbf{CO $\times$ CIB Amplitudes}} \\
	\hline \hline
         Frequency & M23 & K24 & This Work \\
         (GHz) & ($\mu$K$^2$) & ($\mu$K$^2$) & ($\mu$K$^2$) \\
         \hhline{= =|=|=}
         90 & --- & 0.4 -- 1.1  & 0.34 -- 1.08 \\
         150 & 0.89  & 0.8 -- 2.3  & 0.55 -- 1.76 \\
         220 & 3.20  & 3.1 -- 7.2  & 1.07 -- 5.56  \\
    \end{tabular}
    \caption{Amplitudes $\mathcal{D}_\ell$ of the \cross cross-spectra at $\ell=3000$ for common CMB observing frequencies. We compare the results of this work to those of M23 and K24. Note that these are double the \cross spectra as this is the contribution to the total observed power spectrum (see Eq.~\eqref{eq:ps_single_nu} for details). Our range of \cross are consistent with those from previous works.}
    \label{tab:coxcib}
\end{table}

We find that while the amplitude of the CO auto-spectra rises at higher multipoles, they are around one order of magnitude lower than the expected kSZ power spectrum (also shown in Fig.~\ref{fig:dls}) at all $\ell$ for each of our frequencies. Analyses involving higher-order statistics, like CMB lensing, may be affected by higher-point statistics of the CO fluctuations.

Our CO auto-spectra are generally lower than those from M23 and K24 (see Table~\ref{tab:co_auto}) but of the same order of magnitude at the upper end of our range. The auto-spectrum from M23 is comparable to the highest auto-spectrum in our range at 150 GHz but about half an order of magnitude higher at 220 GHz. Similarly, K24 predict higher CO auto-spectra than ours at each frequency, although our ranges have a small overlap at each frequency. Both M23 and K24 found the CO auto-spectrum to exceed that of the kSZ at 220 GHz for $\ell>7000$ and $\ell>8000$, respectively. In contrast, we find the CO auto-spectrum to be below the kSZ auto-spectrum at all considered multipoles and frequencies. 

The \cross cross-correlation at a single frequency shown in Fig.~\ref{fig:dls} includes both cross terms of \cross to capture the entire contribution of the cross-correlation to the total observed power spectrum (see Eq.~\eqref{eq:ps_single_nu} for details). The range of CO models we consider shows a one order of magnitude spread in power at all multipoles and at each frequency. While the upper end of our range becomes comparable to the expected kSZ at only the largest considered multipole of $\ell = 10000$ at 90 GHz, it does so at $\ell\sim5000$ at 150 GHz, and at 220 GHz, \cross is at least comparable to the kSZ auto-spectrum amplitude at all multipoles. This agrees with the results from M23, which predicts \cross to become comparable to the kSZ auto-spectrum at $\ell\sim6500$ and $\ell\sim2500$ at 150 and 220 GHz, respectively. This suggests that the \cross cross-spectrum may pose a potential bias to kSZ measurements. However, this bias is highly model-dependent. 

Interestingly, while our CO auto-spectra are generally lower than those of previous works, our \cross are in good agreement with both M23 and K24, particularly at lower frequencies (see Table~\ref{tab:coxcib}). At 220 GHz, the range from K24 shifts slightly higher than ours, but they are still in good agreement with each other.

\subsection{Cross-frequency Spectra}

%Frequency Spectra
\begin{figure*}
    \centering
    \includegraphics[width= \textwidth]{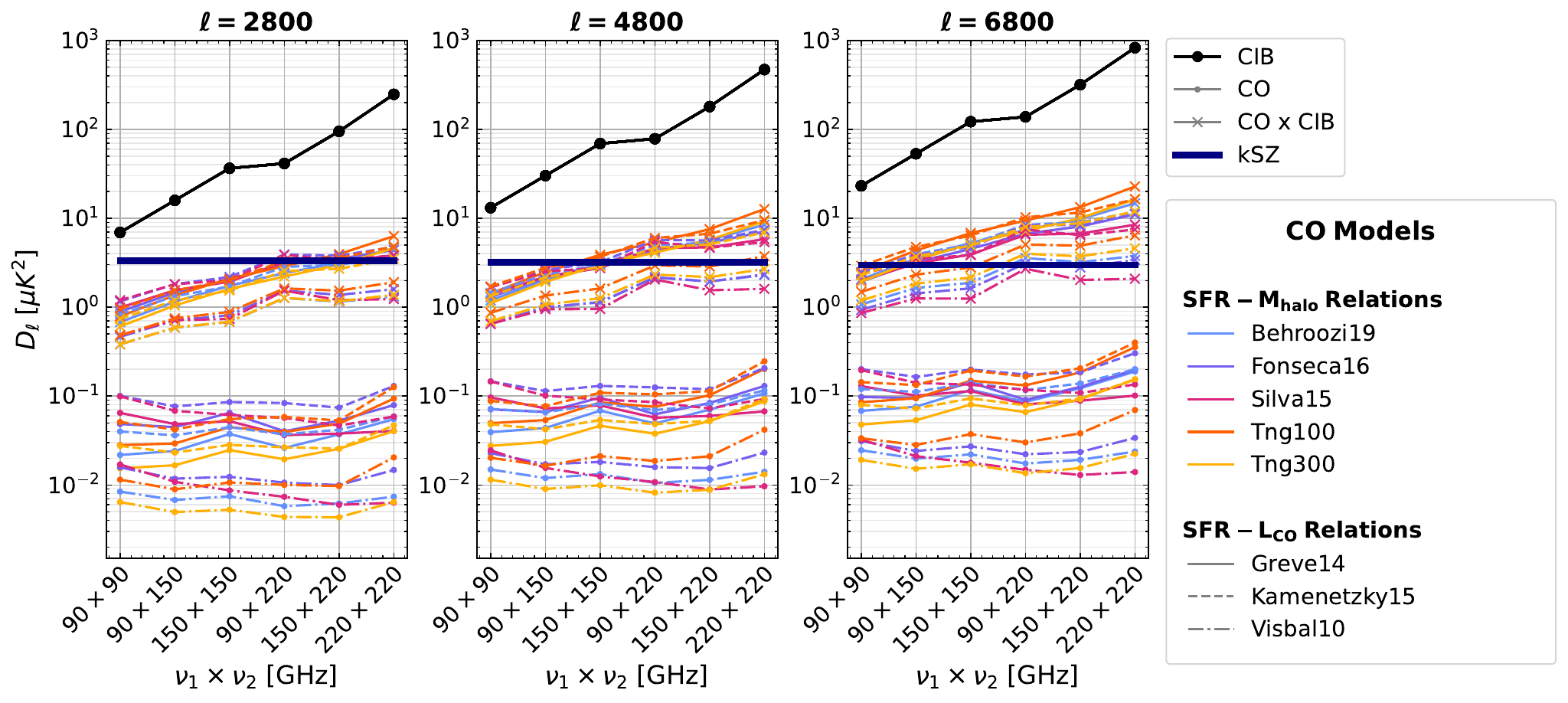}
    \caption{Auto- and cross-spectra of CO, CIB, and \cross for each frequency pair evaluated at several scales $\ell$. The frequency dependence of the CO autospectrum is model-dependent in both amplitude and shape. The \cross spectra are heavily influenced by the CIB spectrum and so monotonically increase with frequency. }
    \label{fig:frequency_templates}
\end{figure*}

We also present cross-frequency spectra for each of our power spectra. Component separation methods, like the internal linear combination (ILC) method, leverage multifrequency cross-correlations between the CMB and astrophysical sources to isolate these signals \citep{WMAP:2003cmr, Planck:2015vgm, Madhavacheril:2019nfz, ACT:2023wcq, SPT-SZ:2021gsa}. Thus, the multifrequency cross-correlations of CO provide additional information that is important for separating extragalactic CO from similar foregrounds, like the CIB.

Our cross-frequency spectra are shown in Figure~\ref{fig:frequency_templates}. We compute these according to Eqs.~\eqref{eq:ps_single_nu} and \eqref{eq:ps_multi_nu}. We find CO autospectrum to vary with frequency between 1 and 2 orders of magnitude, depending on the model. The range of CO autospectra is relatively constant at lower frequencies and increases modestly at 220 GHz. The frequency dependence is not purely a function of the \masshalo -- SFR or scaling relations. For instance, the \verb|Silva15|, \verb|Visbal10| model decreases monotonically with increasing frequency between 90 GHz and 150 GHz, whereas the \verb|Silva15|, \verb|Greve14| model decreases and then increases over the same frequency range. While some models predict a spectrum relatively insensitive to frequency, and therefore degenerate with that from kSZ, the amplitude of the highest predicted CO model is still an order of magnitude lower than the expected kSZ signal. 

Our computed \cross cross-frequency spectra are also shown in Fig.~\ref{fig:frequency_templates}. The range of predicted spectra increases with frequency, particularly at 220 GHz: the \cross spectra at 90 GHz span 0.5 orders of magnitude but 1 order of magnitude at 220 GHz. The spectra themselves generally increase with frequency for all models, and all of the models have similar frequency dependencies. As a result, there is limited degeneracy between the shapes of the kSZ auto-spectrum and \cross frequency spectra. Because of this, multifrequency analyses may be able to separate the kSZ signal from that of \cross, which is a potentially significant bias on kSZ measurements from extragalactic CO, despite their comparable amplitudes.

The relatively similar cross-frequency spectra for each model, particularly the \cross spectra, indicate that a small number of templates for each signal may be sufficient to capture their frequency dependence. We explore this in Section~\ref{sec:PCA}.

\subsection{Correlation Coefficients}

%Correlation Coefficients
\begin{figure*}
    \centering
    \includegraphics[width= \textwidth]{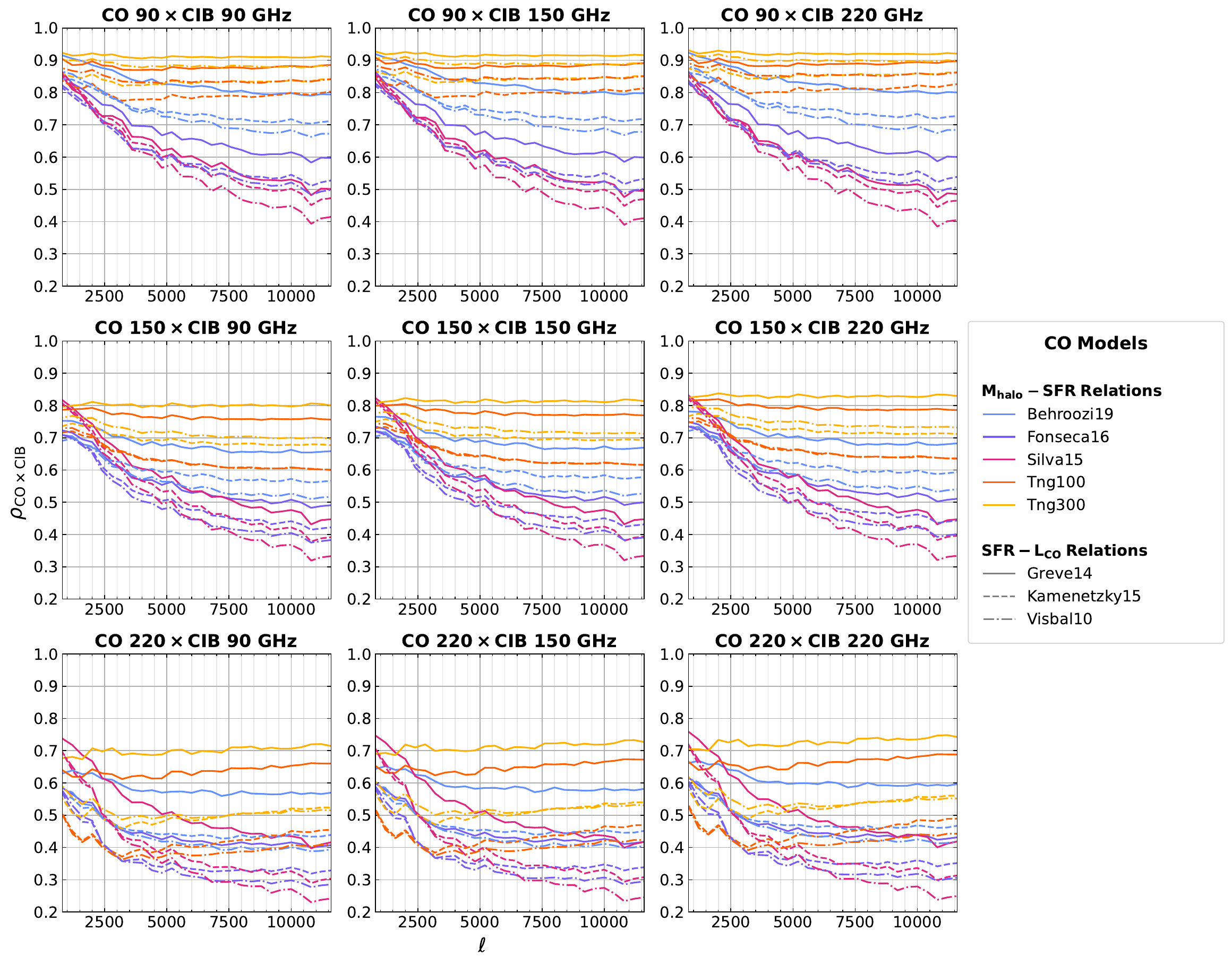}
    \caption{The cross-frequency correlation coefficients for \cross as a function of $\ell$. Different CO models predict differences in both the amplitude and $\ell$ dependence of the correlations. The \cross decoherence is caused almost entirely by the CO decoherence between different frequencies, suggesting that ILC may not be ideal for isolating CO. Given the large range of correlations from the model choices, a multi-frequency power spectrum analysis may be able to distinguish between the models. } 
    \label{fig:rho}
\end{figure*} 

Component separation methods like ILC are most effective when the signals of interest maintain their spatial correlation across frequencies. Reductions in these correlations across frequencies --- so-called ``frequency decoherence" --- result in noise in these types of component separation methods. In contrast, when modeling the foregrounds in a power spectrum analysis, a high degree of correlation between CO and an explicitly modeled foreground, like the CIB, can help remove the CO when accounting for the CIB.

We quantify the decoherence with the correlation coefficient $\rho$ between our simulated CO and CIB for each CO model, which is defined as
\begin{equation} \label{eq:rho}
    \rho_{\nu1\nu2}^{A \times B} = \frac{C_{\ell}^{\mathrm{A_{\nu1}\times B_{\nu2}}}}{\sqrt{C_{\ell}^{\mathrm{A}_{\nu1}\times \mathrm{A}_{\nu1}}  C_{\ell}^{\mathrm{B}_{\nu2}\times \mathrm{B}_{\nu2}}}}. 
\end{equation}
A and B are labels for the signals of interest (CO and CIB in this case), and $\nu1$ and $\nu2$ are the observing frequencies for each signal. We only consider \cross as we find this to be the dominant effect of extragalactic CO in CMB surveys. 

These cross-frequency correlation coefficients are shown in Fig.~\ref{fig:rho}. CO and the CIB are highly correlated in the clustering regime at lower $\ell$, particularly at lower frequencies: for example, CO 90 GHz $\times$ CIB 90 GHz are $\approx83\% - 93\%$ correlated. While some models predict a constant correlation across all scales, others predict a significant reduction in the correlation on smaller scales. This $\ell$-dependence appears to be largely --- but not entirely --- caused by the choice of \masshalo -- SFR relation, while the SFR -- $L_\mathrm{CO}$ relation largely --- but, again, not entirely --- sets the amplitude of the correlation. This indicates that while the S12 model for the CIB does not exactly correspond to one of the \masshalo -- SFR relations for CO that we consider, we would still obtain a significant variation in the \cross correlation based on different SFR -- $L_\mathrm{CO}$ modeling choices. Because of this, these differences in correlations may be used to distinguish between CO models.

We find that the \cross correlation decreases with increasing frequency. While the range of our considered CO models decreases in correlation with the CIB by about 20\% from 90 to 220 GHz, the reduction in correlation for a given model can exceed this; for instance, the \cross correlation for the \verb|TNG100|, \verb|Kamenetzky15| CO model reduces by $> 30\%$ across this frequency range. Given the different redshifts that contribute to the CO power at each frequency channel (see Fig.~\ref{fig:dcdz}), this is not surprising.  We also find that this correlation is almost entirely set by the CO frequency and is insensitive to that of the CIB. The CIB itself decoheres by $\sim5\%$ for adjacent frequency channels \citep{Planck:2013wqd, stein_2020_WebskyExtragalacticCMB, omori_2022_AgoraMultiComponentSimulation}, so we expect the CIB frequency decoherence to be a much smaller effect than that of CO. Because of this, most of the \cross decoherence is due to that of CO.

This large frequency decoherence suggests that the CO might best be treated as separate signals with unique spatial positions at each frequency. This stems from the fact that considerably different redshift ranges contribute to each frequency (see Fig.~\ref{fig:dcdz}). Therefore, isolating extragalactic CO may not be amenable to ILC methods. If one is only concerned about the removal of CO, like in CMB power spectrum analyses targeting the CMB or kSZ autospectrum, then the high spatial correlation predicted by the upper end of our range suggests that CO may potentially already be significantly removed in these analyses at low frequencies. However, given the wide range of predicted correlations and the lower correlations at higher frequencies, the most optimal approach would be to marginalize over all possible \cross models in such analyses. One method for doing so is with a PCA-based parameterization for the \cross power, which we explore in Section~\ref{sec:PCA}.

\section{Biases to Parameters from CO} \label{sec:parameterbiases}

Previous small-scale CMB power spectrum analyses for ground-based CMB surveys (e.g., with ACT \citep{choi_2020_AtacamaCosmologyTelescope,ACT:2025fju} and SPT \citep{George:2014oba,SPT:2020psp,balkenhol_2023_MeasurementCMBTemperature}) have not typically included CO, or its correlation with other tracers, among the foregrounds that are normally included in parameter fitting. 
Instead, the extragalactic components of these fits have typically included the power spectra of the tSZ and kSZ; the CIB, including spatially clustered and unclustered (Poisson) components; and the spatial correlations between these, typically including only those between the CIB and tSZ.  We have shown that, as with M23 and K24, the \cross correlation can potentially be as large as some of these components, so in this section, we compute the degree to which the parameters that describe other components of the foreground model would be biased if it is ignored.

Here, we consider a simplified version of such foreground models.  Specifically, we use the model implemented in the Simple Internal Linear Combination (SILC) code\footnote{SILC code: \url{https://github.com/nbatta/SILC}}, which was originally used for forecasting constraints on cosmological Rayleigh scattering by \citet{Zhu:2022dqi} and is broadly based on the parameterization from \citet{dunkley_2013_AtacamaCosmologyTelescope}.   Specifically, we include tSZ and kSZ power, with scale-dependent templates from \citet{Battaglia:2010tm}, and with amplitudes given by the parameters $a_\mathrm{tSZ}$ and $a_\mathrm{kSZ}$ respectively; unclustered radio galaxies, with an amplitude parameter $a_\mathrm{radio}$ and a power-law spectral index parameter $\alpha_\mathrm{radio}$; clustered and Poisson components of the CIB, with amplitudes  $a_\mathrm{c}$ and $a_\mathrm{p}$, and a common spectral index $\alpha_\mathrm{CIB}$; and a template for the  CIB$\times$tSZ correlation \citep[e.g.,][]{Addison:2012my} represented by a separate amplitude parameter $a_{\mathrm{CIB}\times\mathrm{tSZ}}$.  To focus on the harmonic modes that are most important for the foreground model parameters rather than those that describe the (lensed) primordial CMB, we restrict our study to $\ell > 3000$.  To simplify the treatment, we include flexibility for the CMB itself on these scales by varying only a single amplitude parameter, $a_\mathrm{CMB}$.  We note that on most of these scales, the CMB power spectrum is dominated by the lensing effect \citep[e.g.,][]{Lewis:2006fu} and that, at least within the $\Lambda$CDM model, its properties are known to much higher precision than the foreground components.  Our baseline model for the CMB and foregrounds on these scales thus includes 9 free parameters $\{\lambda_i\}$: seven that control the amplitudes of components at all frequencies,  and two that control the spectral indices for the emissive sources.  Critically, in our baseline model of the CMB and foregrounds, we do \emph{not} include a model for any CO components: our goal is to determine the impact on the other parameters if CO is completely ignored in an analysis of survey data.

For the survey noise properties, we consider two representative surveys: an ACT-like survey, based roughly on the DR6 parameters presented by \citep{ACT:2025fju}, and a three-frequency version of a survey at CMB-S4-like sensitivity \citep{Abazajian:2019eic}.  Specifically, we assume the ACT-like dataset to have white-noise levels of $[15.0,18.2,82]\,\mu$K-arcmin at $[90,150,220]$\,GHz respectively; while the real ACT maps have spatially anisotropic noise levels \citep{ACT:2025xdm}, these are the approximate effective white noise levels of the DR6 power spectra \citep{ACT:2025fju}.   We take the beams to have full widths at half-maxima of $[2.37,1.50,1.02]$\,arcmin, and for the sky area to be 11,000 square degrees.  For the CMB-S4-like survey, we assume white-noise levels of $[1.3,1.8,9.1]\,\mu$K-arcmin, the same resolutions as the ACT-like survey, and a sky area of 17,000 square degrees. 

\begin{figure}[t]
    \centering
    \includegraphics[width=\columnwidth]{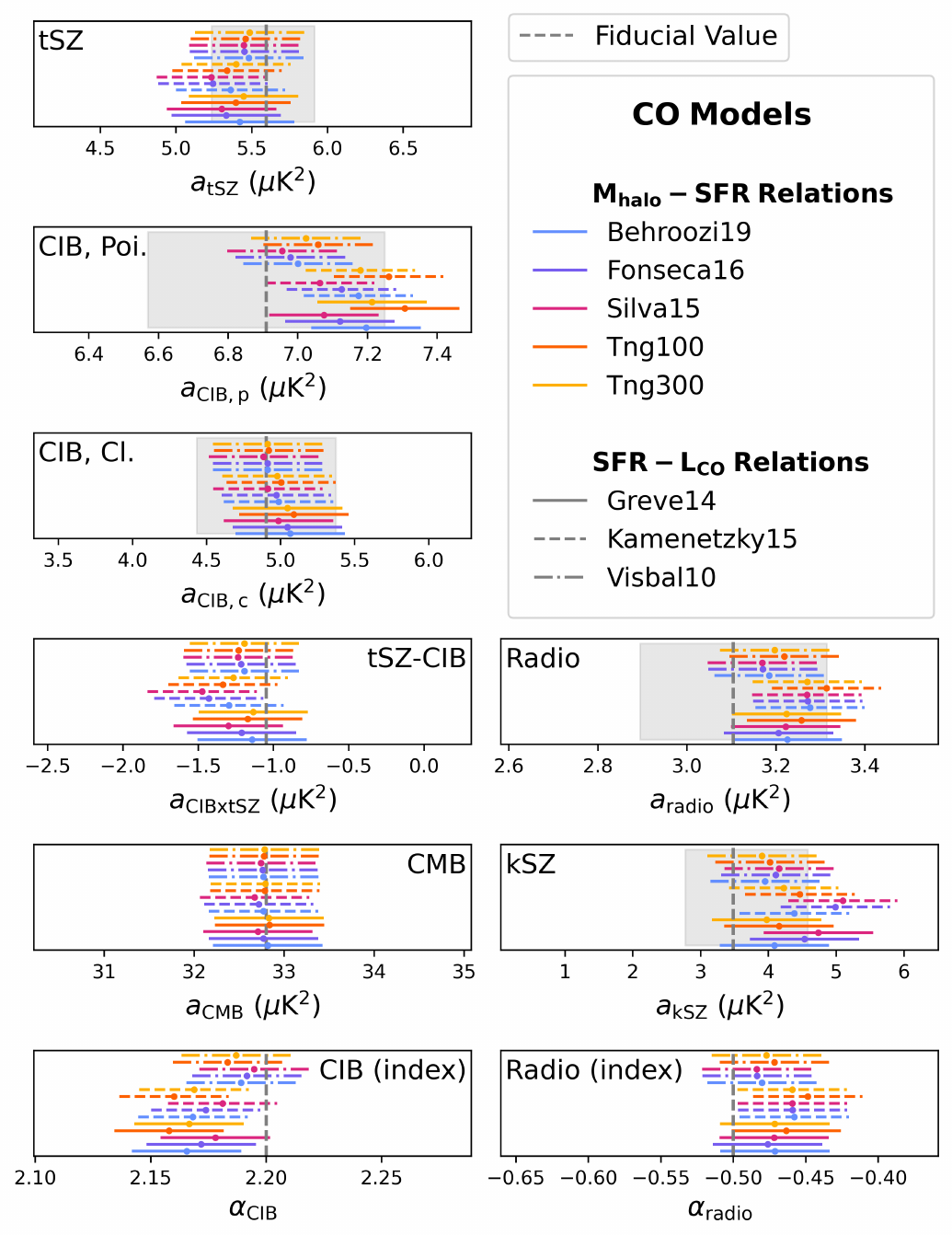}
    \caption{Forecasted parameter shifts on other CMB foreground parameters if the \cross power spectra are not included in the model, given the noise properties of the ACT-like survey.  Each panel corresponds to a parameter that we allow to vary within our foreground model, with fiducial values given by the grey dashed line.  Each \cross model (colored whiskers) gives a unique expected bias to these parameters, which are as large as or larger than the statistical uncertainty for some CO models and some parameters.  Shown for comparison are some of the uncertainties from the real ACT DR6 analysis, centered here on the fiducial values   \citep[light grey shading:][]{ACT:2025fju,Beringue:2025bur}; these statistical uncertainties are comparable to our forecasts, but larger in some cases.   } 
    \label{fig:param_whisker_act}
\end{figure}

Given these ingredients, we form the $3\times3$ matrix of total observed power spectra $(\mathbf{C}_{\ell})_{ab} = C_\ell^{\mathrm{tot}_{\nu_a} \times \mathrm{tot}_{\nu_b}}$; the diagonals of this matrix include the assumed noise power for the given survey.  We then compute the Fisher information matrix between parameters $\lambda_i$ and $\lambda_j$, according to \citep[e.g.,][]{Tegmark:1996qt,Zaldarriaga:1997ch,Hamimeche:2008ai,2010LNP...800..147V} 
\begin{equation}
F_{ij} = \sum_\ell \frac{2\ell + 1}{2} f_\mathrm{sky} \mathrm{Tr} \left(\mathbf{C}_\ell^{-1} \frac{\partial \mathbf{C}_\ell}{\partial \lambda_i} \mathbf{C}_\ell^{-1} \frac{\partial \mathbf{C}_\ell}{\partial \lambda_j} \right)  ,
\label{eq:fishermatrix}
\end{equation}
where $f_\mathrm{sky}$ is the sky fraction of the survey.

For this demonstrative exercise, we restrict the sums to be within $3000 \leq \ell \leq 8000$; a more involved analysis would extend to the lower end of the multipole range and would include more parameters to describe the lensed CMB power spectrum.  The parameter covariance matrix we obtain by inverting this Fisher matrix indicates that all nine parameters can be constrained with the survey properties we assumed, though some parameters are mildly degenerate, as expected.  Parameter uncertainties are broadly consistent with previous results from relevant surveys \citep{ACT:2025fju,Abazajian:2019eic}.  

Given the Fisher matrix, we then compute the bias to the parameters in our model in the presence of a ``biasing spectrum" $\Delta C_\ell^\mathrm{bias}$  \citep{Huterer:2004tr,LoVerde:2006cj} following
\begin{align}
\Delta \lambda_{k}^\mathrm{bias} = & \sum_i (F^{-1})_{ki} \sum_\ell \frac{2\ell + 1}{2} f_\mathrm{sky} \\ \nonumber & \ \times \mathrm{Tr} \left(\mathbf{C}_\ell^{-1} \frac{\partial \mathbf{C}_\ell}{\partial \lambda_i} \mathbf{C}_\ell^{-1} \Delta \mathbf{C}_\ell^\mathrm{bias}  \right) \ .
\label{eq:fisherbias}
\end{align}

We evaluate this expression with $\Delta \mathbf{C}_\ell^\mathrm{bias}$ set to each power spectrum model of \cross that we have obtained and again sum over the range $3000 \leq \ell \leq 8000$.  We show the parameter bias results for the ACT-like survey as one-dimensional whisker plots in Fig.~\ref{fig:param_whisker_act} and as parameter confidence contours in Fig.~\ref{fig:param_ellipses_act}\footnote{For Figs.~\ref{fig:param_ellipses_act} and \ref{fig:param_ellipses_s4}, we use elements of the \texttt{FisherLens} code developed for \citet{Hotinli:2021umk}: \url{https://github.com/ctrendafilova/FisherLens}}.  We find a range of potential levels of biases; in particular, some parameters in the  ACT-like survey are biased by amounts ranging between negligible and significant fractions of the forecasted statistical uncertainty.  We also show in Fig.~\ref{fig:param_whisker_act} the results for the ACT DR6 measurements of a selection of these parameters from \citet{ACT:2025fju}; with our forecasts, we obtain comparable uncertainties, but which are smaller in some cases than those from the full analysis.  The computed shifts are mainly of the order of or smaller than the statistical uncertainty for ACT DR6, in particular for the parameters that describe the CIB, the radio galaxies, and the kSZ.  This generally agrees with the recent assessment using the K24 templates with the real DR6 likelihoods from App.~C of \citet{beringue_2025_AtacamaCosmologyTelescope}.

Our results for the CMB-S4-like survey are shown in Fig.~\ref{fig:param_ellipses_s4}.  Here, our forecasting predicts uncertainties that are up to two orders of magnitude more constraining than those of the ACT-like survey.   Relative to these uncertainties, we predict very large biases on these parameters if \cross is neglected --- of many tens of standard deviations.  Furthermore, the size of the bias is seen to depend strongly on the CO model assumed.  This indicates that CO$\times$CIB will need to be included in analyses of future datasets, and that the modelling will need to be broad enough to encapsulate any associated modelling uncertainty.  In the next section, we assess one potential way to accomplish this, with minimal prior assumptions about the CO modelling.

\begin{figure}[t]
    \centering
    \includegraphics[width=\columnwidth]{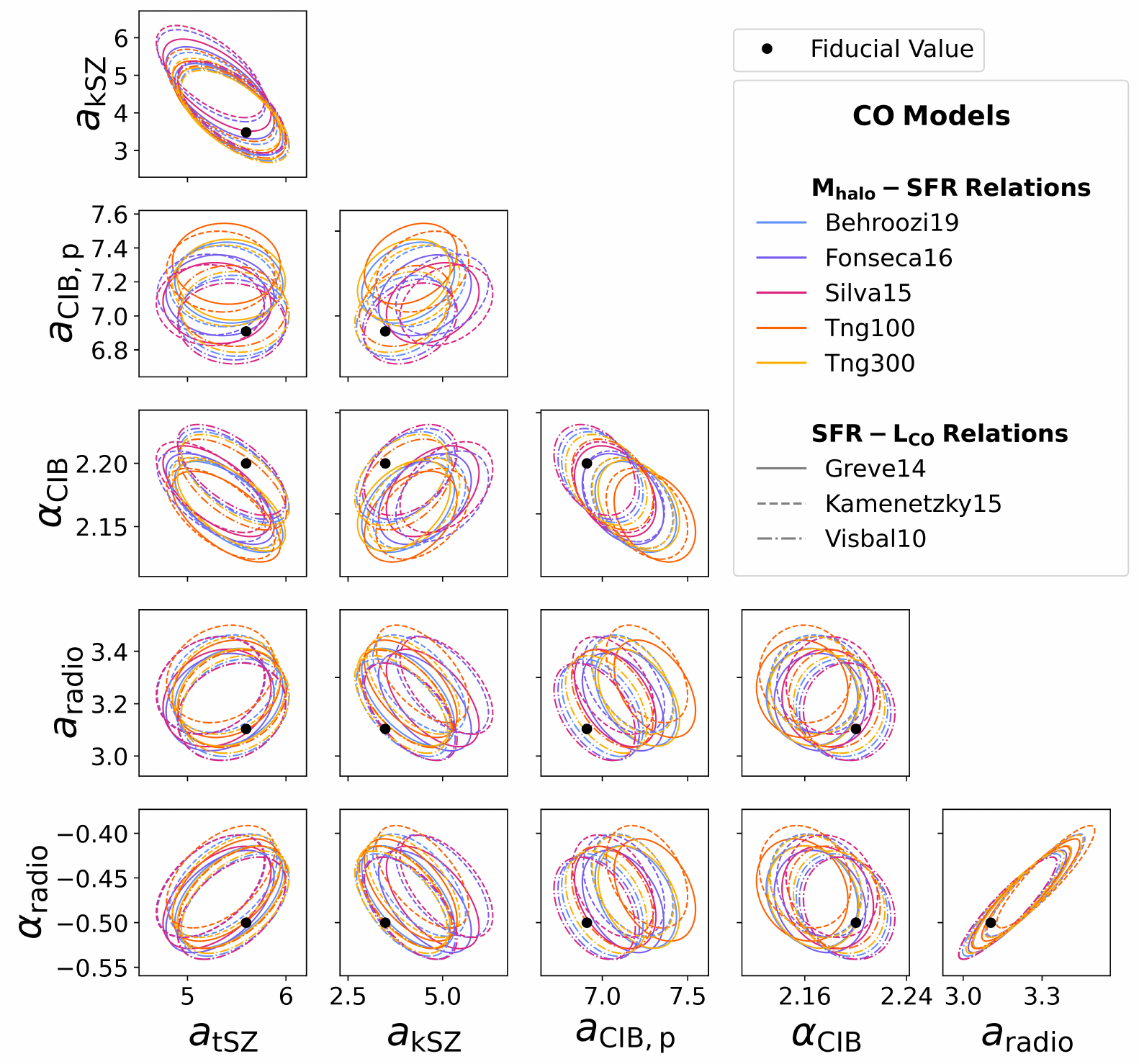}
    \caption{Forecasted shifts in foreground parameter contours if the \cross power spectra are not included in the model for the ACT-like survey. The presence of a given \cross model (colored contours) shifts the parameters away from the fiducial point (black circles).  Note that the parameter uncertainties in our forecast are smaller than for the real ACT DR6 data (see Fig.~\ref{fig:param_whisker_act}).} 
    \label{fig:param_ellipses_act}
\end{figure}

\begin{figure}[t]
    \centering
    \includegraphics[width=\columnwidth]{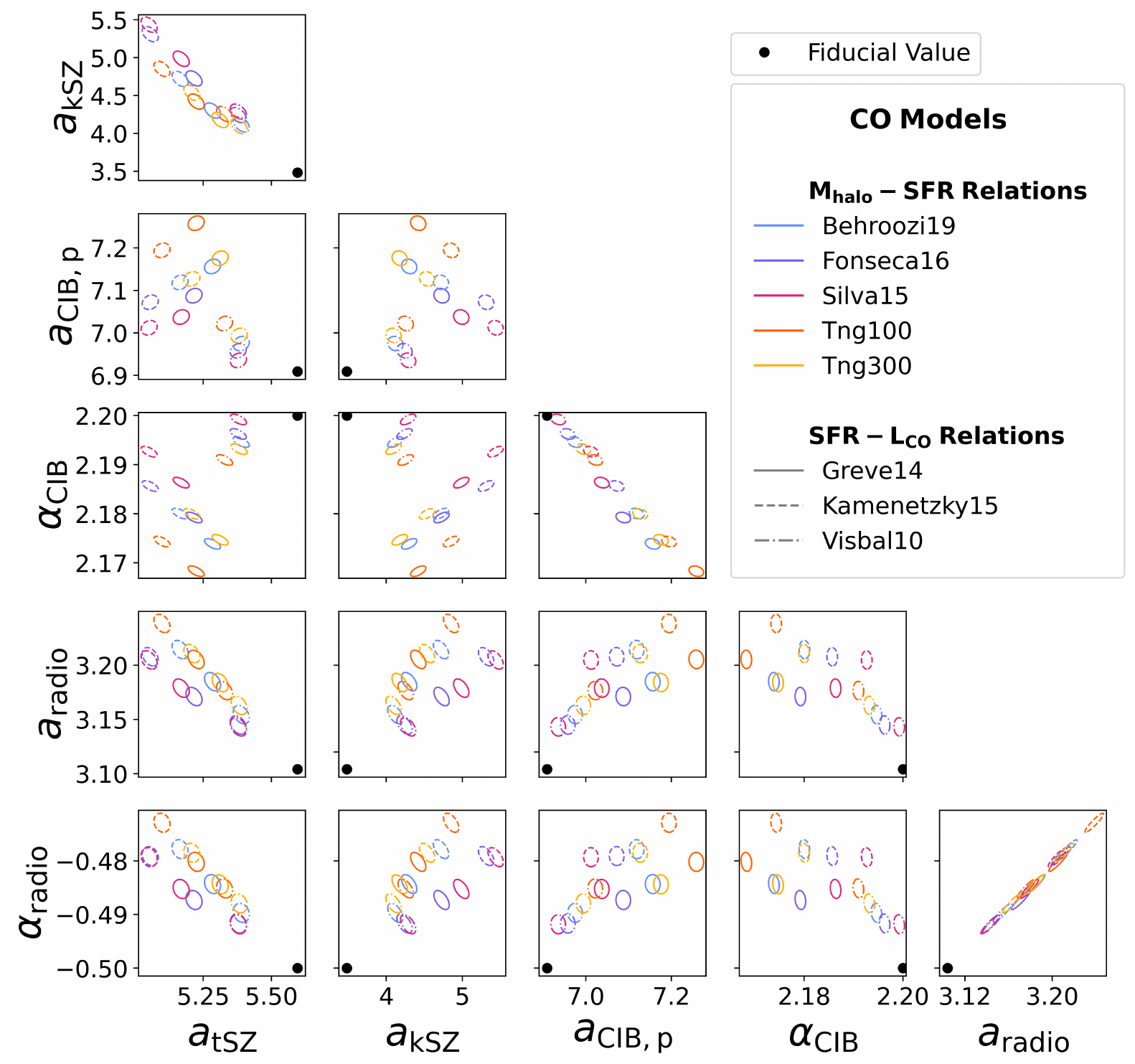}
    \caption{Forecasted parameter shifts on other CMB foreground parameters if the \cross power spectra are not included in the model, for the noise properties of the CMB-S4-like survey, as in Fig.~\ref{fig:param_ellipses_act}; very large biases are seen, meaning \cross must be included in the modeling. The variation in the bias for several parameters, like $a_\mathrm{kSZ}$, demonstrates the need for accurate CO modeling for marginalization with future experiments. } 
    \label{fig:param_ellipses_s4}
\end{figure}

\section{Principal Component Analysis of CO Models} \label{sec:PCA}

To incorporate the CO contribution when interpreting measured CMB power spectra, one would include $C_\ell^{{\rm CO}\times{\rm CIB}}$ and $C_\ell^{{\rm CO}\times{\rm CO}}$ in the model for CMB temperature and either marginalize over or attempt to constrain the parameters of the underlying CO model. The simulation-based predictions presented here and in previous works (M23 and K24) are too cumbersome to re-evaluate during parameter inference due to the large parameter set that describes them. A simpler alternative is to decompose a set of models into orthogonal principal components (PCs) and include a fixed number of these principal components in the total CMB temperature model, each with an amplitude that contributes an additional free parameter. Assuming that the true behavior of CO is within the space spanned by these models, this provides an economical way to account for the uncertainties in CO modeling. Similar approaches have been used for baryonic effects in cosmic shear (e.g.,~\citealt{Eifler:2014iva,Huang:2018wpy}) and modified gravity models in $3\times 2$-pt analyses \citep{Zanoletti:2025xdc}.

We refer to a pair of choices for the $M_{\rm halo}$ -- SFR relation and SFR -- $L_{\rm CO}$ relation as a single ``model," meaning we have $N_{\rm models}=5\times 3 = 15$. For brevity, we also use $F$ to denote a pair of frequency bands that enter the power spectrum (e.g., $90\times150\,{\rm GHz}$), and take a single $F$ to include both permutations of CO and CIB cross-correlations (e.g.,~${\rm CO}\,90\times {\rm CIB}\,150\,{\rm GHz} \;+\; {\rm CIB}\,90\times {\rm CO}\,150\,{\rm GHz}$). Since we consider 3 bands, we have $N_F=6$.

To decompose our set of models into principal components, we form a matrix $\boldsymbol{M}$ of angular power spectra\footnote{We use $\mathcal{D}_\ell^F \equiv \ell(\ell + 1) C_\ell^F/2\pi$ instead of $C_\ell^F$ to reduce the dynamic range spanned by the spectra.}, where each column is a different model $m$ and each row is a different tuple of frequency band pair $F$ and multipole $\ell$:
\begin{equation}
\boldsymbol{M} = 
\left[
\begin{array}{ccc}
\mathcal{D}_{\ell_{\rm min}}^{F_1, m_1} & \mathcal{D}_{\ell_{\rm min}}^{F_1, m_2} & \cdots \\
\vdots & \vdots & \cdots \\
\mathcal{D}_{\ell_{\rm max}}^{F_1, m_1} & \mathcal{D}_{\ell_{\rm max}}^{F_1, m_2} & \cdots \\
\mathcal{D}_{\ell_{\rm min}}^{F_2, m_1} & \mathcal{D}_{\ell_{\rm min}}^{F_2, m_2} & \cdots \\
\vdots & \vdots & \cdots \\
\mathcal{D}_{\ell_{\rm max}}^{F_2, m_1} & \mathcal{D}_{\ell_{\rm max}}^{F_2, m_2} & \cdots \\
\vdots & \vdots & \ddots
\end{array}
\right]\ .
\label{eq:Mmatrix}
\end{equation}
We will restrict our analysis to the CO$\times$CIB cross spectrum, since it dominates over CO$\times$CO, but in principle one could include both types of spectrum in~$\boldsymbol{M}$. The matrix
\begin{equation}
\boldsymbol{C} 
	= \frac{1}{{N_{\rm models}-1}}
	\boldsymbol{M} \boldsymbol{M}^{\rm T}
\end{equation}
is the covariance of $\mathcal{D}_\ell^F$ over our set of models, and the eigenvectors of this matrix correspond to the PCs of these models. In practice, we compute these eigenvectors by performing a singular value decomposition (SVD) of~$\boldsymbol{M}$,
\begin{equation}
\label{eq:svd}
\boldsymbol{M} = \boldsymbol{U} \boldsymbol{\Sigma} \boldsymbol{V}^{\rm T}\ ,
\end{equation}
using the fact that the eigenvectors of~$\boldsymbol{C}$ are precisely the columns of~$\boldsymbol{U}$ (i.e.\ the left singular vectors of~$\boldsymbol{M}$).
Writing the $(F,\ell)$ element of the $i^{\rm th}$ PC as~$U_{(F,\ell),i}$, and its corresponding singular value as $s_i$, any model within our set can be represented as a sum of these PCs with appropriate order-unity coefficients $\alpha_i$:
\begin{equation}
\mathcal{D}_\ell^{F,m} = \sum_{i=1}^{N_{\rm models}} \alpha_i s_i U_{(F,\ell),i}\ .
\label{eq:modesum}
\end{equation}
In Eq.~\eqref{eq:svd}, $\boldsymbol{\Sigma}$ is a diagonal matrix containing the singular values $s_i$, and the columns of $\boldsymbol{V}$ contain the specific values of the $\alpha_i$ coefficients for each model.

\begin{figure*}[t]
    \centering
    \includegraphics[width=\textwidth, trim=7 10 5 0]{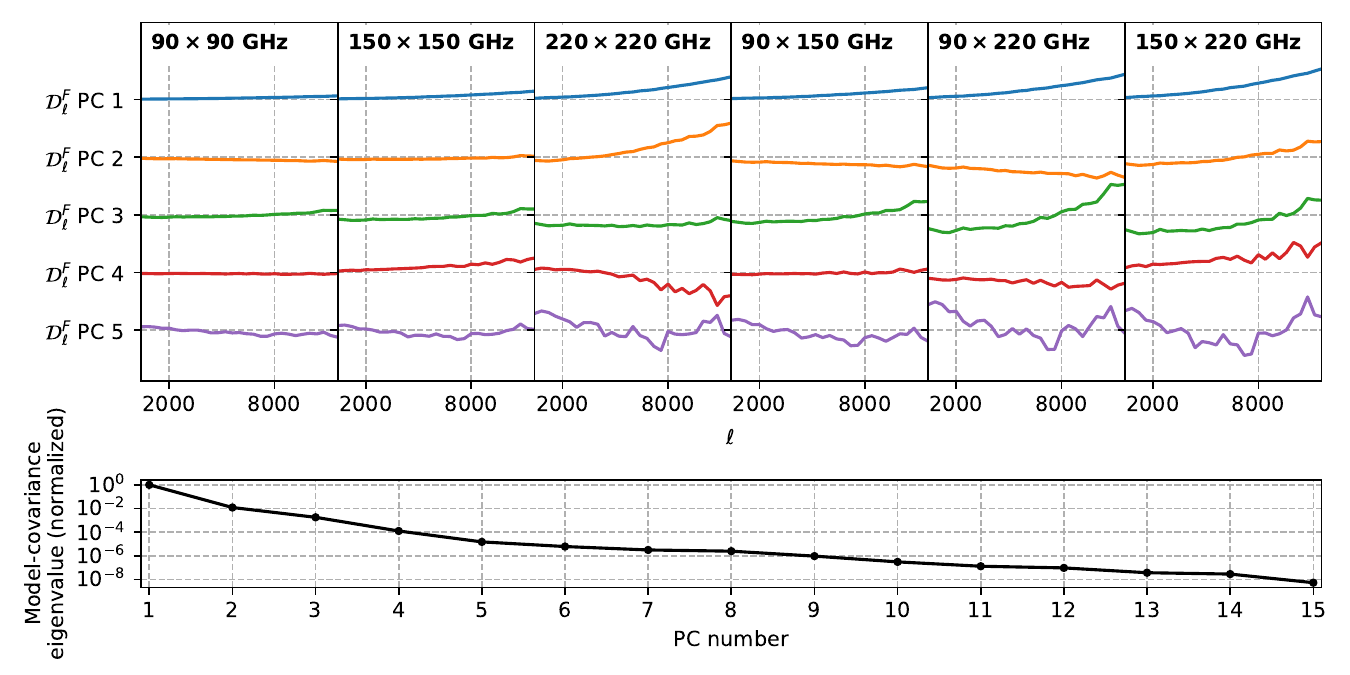}
    \caption{\textit{Upper panel:} First 5 principal components of the set of CO$\times$CIB angular cross power spectra computed from the models explored in this work. The dominant principal component (``PC 1") represents the ``average shape" of $\mathcal{D}_\ell^F$ across all models (where $F$ denotes a pair of frequency bands), while subleading PCs indicate different variations of the models around this average shape, ordered by importance.
    \textit{Lower panel:} Eigenvalues of the covariance of $\mathcal{D}_\ell^F$ across the set of models, normalized to the largest eigenvalue. The steep eigenvalue spectrum indicates that uncertainties in CO modeling can be reduced to uncertainties on coefficients of a small number of PCs.} 
    \label{fig:svd_Dl}
\end{figure*}

In Figure~\ref{fig:svd_Dl}, we show the first 5 PCs, along with the corresponding eigenvalues of $\boldsymbol{C}$ (equal to the squares of the singular values $s_i$), normalized to the largest eigenvalue. The leading PC effectively represents the ``average shape" of $\mathcal{D}_\ell^F$ for CO$\times$CIB across all models, while subleading PCs are orthogonal modes of variation around this shape. For example, the second PC reflects the fact that the slopes of the models (as a function of~$\ell$) at $220\times220\,{\rm GHz}$ show more variation than at the other frequency pairs. 
The partially-jagged appearance of the higher PCs indicates that they are at least partially capturing variations between models due to numerical noise in the simulation measurements, rather than physically-meaningful changes in the spectra. The 6th and higher PCs (not shown) are dominated by these noise variations, which indicates that these PCs should not be included in data analysis or forecasts, unless they are re-computed with using more precise simulation measurements. Furthermore, the steep slope of the eigenvalue spectrum shows that even some of the well-resolved PCs 
can likely be neglected in a practical application. 

One can re-weigh the model spectra in various ways before forming $\boldsymbol{M}$ in order to obtain a set of PCs that is optimized for a specific purpose. For instance, by multiplying each column of Eq.~\eqref{eq:Mmatrix} by the inverse of the Cholesky decomposition of the covariance of $\mathcal{D}_\ell^F$ from a given experiment, one can obtain signal-to-noise PCs, whose ordering is based on the detectability of each successive mode of variation \citep{Huang:2018wpy}. One can then transform these modes from signal-to-noise space back to signal space, and assess
how many of these modes need to be retained in Eq.~\eqref{eq:modesum} to provide sufficient accuracy for the interpretation of the corresponding experiment.\footnote{One can also perform a weighted PCA directly in signal space, by iteratively solving a weighted least-squares problem for each successive PC (e.g.,\ \citealt{COMAP:2024iqp}). We have separately implemented this procedure, and found the results to be identical to the method described in the main text.}

\begin{figure*}[t]
    \centering
    \includegraphics[width=\textwidth, trim=7 10 5 0]{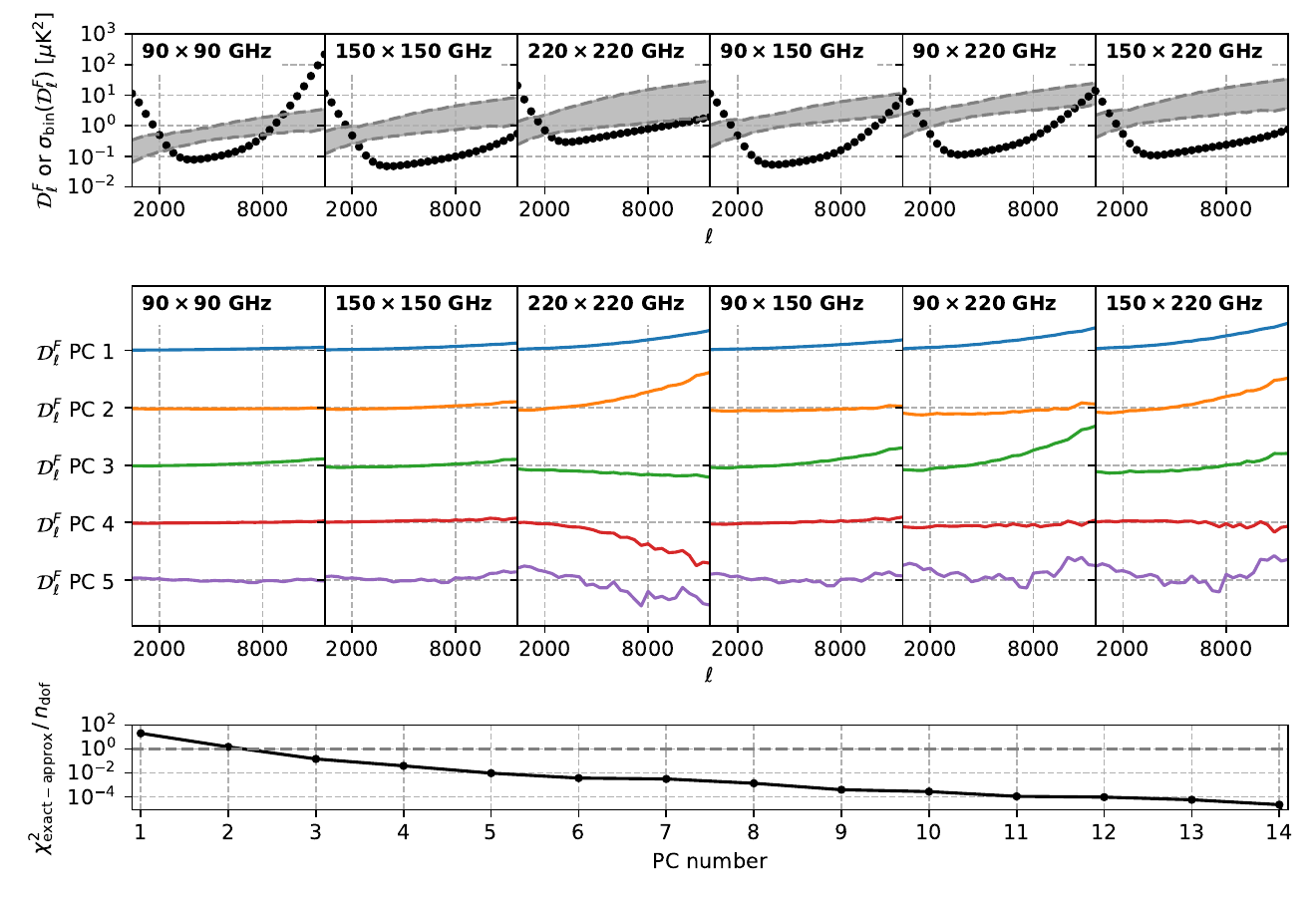}
    \caption{\textit{Upper panel:} The black points show the estimated uncertainties on $\mathcal{D}_\ell^\nu$ from a CMB-S4-like experiment (see main text for details), while the grey bands show the ranges of CO$\times$CIB cross spectra from the models in Sec.~\ref{sec:Results-Discussion}.
    \textit{Middle panel:} First 5 principal components of the signal to noise per bandpower, $\mathcal{D}_\ell^\nu/\sigma_{\rm bin}(\mathcal{D}_\ell^\nu)$. These optimally capture the highest-signal-to-noise variations in $\mathcal{D}_\ell^\nu$.
    \textit{Lower panel:} Reduced $\chi^2$ of a comparison between each simulation measurement of $\mathcal{D}_\ell^{\nu,{\rm CO}\times{\rm CIB}}$ and its $N$-PC approximation. We show the maximum reduced $\chi^2$ over our set of 15 models. Three PCs are required to achieve $\chi^2/n_{\rm dof} \lesssim 1$, indicating that all models are well-described by three PCs within the assumed experimental uncertainties. 
    }
    \label{fig:svd_SN}
\end{figure*}

As an illustration, consider the simple case of a diagonal, Gaussian noise covariance, associated with bandpowers with width $\Delta\ell=400$ (corresponding to the binning that was used in our simulation measurements). In this case, we can obtain signal-to-noise PCs by 
dividing each element of Eq.~\eqref{eq:Mmatrix} by $\sigma_{\rm bin}(\mathcal{D}_\ell^F)$, where
\begin{equation}
\sigma_{\rm bin}(\mathcal{D}_\ell^F) 
	= \left[ \sum_{\ell'=\ell-\Delta\ell/2}^{\ell+\Delta\ell/2} 
		\frac{1}{\sigma(\mathcal{D}_{\ell'}^F)^2} 
	\right]^{-1/2}
\label{eq:sigmabin_Dl}
\end{equation}
and
\begin{align*}
\sigma(\mathcal{D}_\ell^F)^2
	&= \frac{1}{f_{\rm sky}(2\ell+1)} \left( 
		\left[ \mathcal{D}_\ell^{\nu_1 \nu_2,{\rm total}} \right]^2
		\right. \numberthis \label{eq:sigma_Dl} \\
&\quad +
	\left. \left[ \mathcal{D}_\ell^{\nu_1 \nu_1,{\rm total}} 
	+ N_\ell^{\nu_1} \right] \left[ \mathcal{D}_\ell^{\nu_2 \nu_2,{\rm total}} 
	+ N_\ell^{\nu_2} \right]
	\right), 
\end{align*}
and then carrying out the SVD.
In Eq.~\eqref{eq:sigma_Dl}, $\mathcal{D}_\ell^{\nu_1\nu_2,{\rm total}}$ is the total power spectrum (including primary fluctuations, uncleaned foregrounds, etc.) for the cross-correlation of bands $\nu_1$ and $\nu_2$ in pair $F$, $N_\ell^\nu$ is the noise power spectrum in band $\nu$, and $f_{\rm sky}$ is the accessible sky fraction. The resulting PCs represent variations in $\mathcal{D}_\ell^F/\sigma_{\rm bin}(\mathcal{D}_\ell^F)$, and can therefore be converted back to variations in $\mathcal{D}_\ell^F$ by multiplying by~$\sigma_{\rm bin}(\mathcal{D}_\ell^F)$.\footnote{PCs of $\mathcal{D}_\ell^F/\sigma_{\rm bin}(\mathcal{D}_\ell^F)$ will be orthonormal by design, in the sense that $\sum_{F,\ell} U_{(F,\ell),i} U_{(F,\ell),j} = \delta_{ij}$, but multiplying each PC by $\sigma_{\rm bin}(\mathcal{D}_\ell^F)$ generally will not preserve orthonormality. However, once a set of relevant PCs has been identified, the corresponding set of transformed PCs can easily be orthogonalized, e.g.,\ via Gram-Schmidt orthogonalization.}

The upper panels of Figure~\ref{fig:svd_SN} show $\sigma_{\rm bin}(\mathcal{D}_\ell^F)$ for our 6 frequency-band pairs for a three-frequency experiment at CMB-S4-like sensitivity, assuming $f_{\rm sky}=0.4$ and using noise and foreground power spectra from the \texttt{SILC} code, as used in Section~\ref{sec:parameterbiases}. The panels also show shaded bands corresponding to the ranges of CO$\times$CIB cross spectra presented in Sec.~\ref{sec:Results-Discussion}. 
We see that there is significant sensitivity to the CO$\times$CIB signal across a wide range of multipoles in several frequency-band pairs we consider.
The middle panel of Figure~\ref{fig:svd_SN} shows the first 5 PCs of $\mathcal{D}_\ell^F$ obtained from a signal-to-noise PCA. The shapes of these PCs are broadly similar to those in the unweighted case in Figure~\ref{fig:svd_Dl}, but are optimized to capture variations at multipole and frequency ranges with lower noise.

To roughly assess how many CO$\times$CIB PCs should be included in the CMB temperature model when interpreting data from this experiment, we compute a chi-squared statistic comparing a given exact simulation output for $\mathcal{D}_\ell^{F,{\rm CO}\times{\rm CIB}}$ with its approximation by the first $N$ PCs from Figure~\ref{fig:svd_SN}:
\begin{equation}
\chi^2 = \sum_{F,\ell}
	\frac{(\mathcal{D}_\ell^{F,m,N\,\text{PCs}} - \mathcal{D}_\ell^{F,m,{\rm exact}})^2}
	{\sigma_{\rm bin}(\mathcal{D}_\ell^F)^2}\ .
\end{equation}
We take the maximum $\chi^2$ value over the 15 models we consider, and plot the results in the lower panel of Figure~\ref{fig:svd_SN}. If $\chi^2/n_{\rm dof} \lesssim 1$, where $n_{\rm dof} = 174 - N$ (from the 174 $(F,\ell)$ pairs at which we have measured $\mathcal{D}_\ell^F$), then the simulation model is well-described by an $N$-PC approximation within the assumed experimental uncertainty.\footnote{We have compared this heuristic criterion to a more systematic version, in which each $\chi^2$ value is converted into a $p$-value using a $\chi^2$ distribution with $174-N$ degrees of freedom. In this version, we ask how many PCs are needed to achieve $p>0.16$, corresponding an inability to reject the hypothesis that the simulations are well-described an $N$-PC approximation at greater than  $1\sigma$. The associated threshold in $\chi^2/n_{\rm dof}$ values is roughly~$1.1$ over the range of $N$ we consider. Given the steep behavior of $\chi^2/n_{\rm dof}$ with $N$ shown in the lower panel of Figure~\ref{fig:svd_SN}, this criterion leads to conclusions that are identical to those derived from the simpler criterion in the main text.} We find that a 3-PC approximation meets this criterion. For lower-sensitivity experiments like SO or ACT, it's likely that only one or two PCs would be required. 

We note, however, that this is only an approximate method to assess the required number of PCs; a more rigorous assessment would carry out simulated analyses with various numbers of PCs, and quantify how many PCs are required before posteriors on other model parameters acquire significant bias. We leave such a test for future work.

\section{Conclusion} \label{sec: Conclusion}

In this work, we investigated the contributions of extragalactic CO to CMB temperature power spectrum measurements. We applied a range of CO models used in the literature to N-body simulations, and used these simulations to estimate the contribution to the observed CMB temperature power spectrum from the CO auto-spectrum of each model. We also simulated the CIB and computed the cross-correlation between CO and the CIB for each of these CO models. 

We showed that in a broadband CMB survey, extragalactic CO across almost all of cosmic time contributes to the total observed power spectrum. Since CO traces the star formation history of the universe, the majority of the contribution is sourced during cosmic noon ($z\sim2$). We found that the impact of the total CO autospectrum is small for all of our considered CO models. However, we also found that CO may be upwards of $\approx90\%$ correlated with the CIB, since both trace the history of star formation. Because of this, the \cross amplitude may be comparable to the kSZ autospectrum, particularly at higher frequencies. 

These findings are consistent with previous works by M23 and K24. We generally predicted lower CO autospectra than these works, with only mild agreement in the upper end of our considered range. Despite this, our \cross spectra agreed well with both of these works. We leave a detailed comparison between these models and simulations across all three works to a future investigation.

Our work demonstrated that even qualitative conclusions about the impact of extragalactic CO on CMB surveys are most limited by the large theoretical uncertainty of CO modeling, particularly at high redshifts. This uncertainty spans up to 2 orders of magnitude at the power spectrum level within the models we considered. For instance, while we found that the largest potential bias from CO on kSZ measurements will come from its cross-correlations with the CIB, differences in modeling choices alone can change the power of this cross-correlation from greater than the kSZ autospectrum to 10\% of it. This wide range of theoretical models is the largest source of uncertainty when quantifying the effects of extragalactic CO in broadband surveys and must be understood for current and future small-scale CMB experiments.

To further investigate the effect of the \cross cross-correlation on CMB foregrounds, we performed a Fisher calculation with a foreground model similar to those used in current CMB analyses. This allowed us to calculate the bias of each foreground parameter from our simulated \cross spectra that we expect if the \cross correlation is not included in the analysis. We found these biases to be comparable to the uncertainty of these parameters from recent constraints with ACT DR6 data. These biases become significantly large for a future, CMB-S4-like survey, particularly the biases to the tSZ, kSZ, and radio source parameters. Thus, the \cross cross-correlation will likely need to be considered in current foreground analyses, and certainly will need to be for next-generation CMB datasets. 

Despite the considerable theoretical uncertainty, there are some avenues to mitigate the effects of CO in CMB surveys. While the CO frequency decoherence suggests that ILC methods may not cleanly isolate CO, one might still marginalize over a range of CO models in a CMB power spectrum foreground analysis. To this end, we conducted a PCA analysis to capture this wide range of expected \cross power. We found that 3 principal components are sufficient to capture the range of considered CO models for a three-frequency experiment with CMB-S4-like sensitivity. Given the wide variation of predicted \cross spectra amplitudes, $\ell$-dependence, and frequency dependence, the precision of future CMB experiments may be able to place constraints on, or even rule out, some CO models. Thus, while challenging, future CMB experiments will likely offer an opportunity to study extragalactic CO.

\section*{Acknowledgements}

We would like to thank Jos\'{e} Luis Bernal, Nickolas Kokron, Joel Meyers, Allison Noble, and Alex Pigarelli for helpful discussion.  

We acknowledge the indigenous peoples of Arizona, including the Akimel O’odham (Pima) and Pee Posh (Maricopa) Indian Communities, whose care and keeping of the land has enabled us to be at ASU’s Tempe campus in the Salt River Valley, where this work was conducted.

YM and AvE were supported by NASA grants 80NSSC23K0747, 80NSSC24K0665, and NSF AAG grant 588167.  AvE and SF were supported by NASA grant 80NSSC23K0464. AR acknowledges support from NASA under award number 80NSSC18K101493.

Computations were performed on the Niagara supercomputer \citep{ponce_2019_DeployingTop100Supercomputer} at the SciNet HPC Consortium \citep{loken_2010_SciNetLessonsLearned}. SciNet is funded by Innovation, Science and Economic Development Canada; the Digital Research Alliance of Canada; the Ontario Research Fund: Research Excellence; and the University of Toronto.

% References %
\bibliographystyle{aasjournal}
% \bibliography{refs}
\bibliography{refs,refs2}

\end{document}